\documentclass{aa}

\usepackage{epsf}

\begin{document}

\thesaurus{07(03.13.3; 08.01.3; 08.02.1)}

\title{The use of the {\sc NextGen} model atmospheres for cool giants in a 
light curve synthesis code}

\author{J. A. Orosz\inst{1} \and P. H. Hauschildt\inst{2}}

\offprints{J. A. Orosz}

\institute{Astronomical Institute, Utrecht University, PO Box 80000,
        NL-3508 TA Utrecht, The Netherlands, J.A.Orosz@astro.uu.nl
        \and
        Department of Physics \& Astronomy and 
Center for Simulational Physics, The University of Georgia,
Athens, GA 30602-2451, USA, yeti@hobbes.physast.uga.edu}

\date{received 3 May 2000; accepted 29 September 2000}

\titlerunning{The use of {\sc NextGen} model atmospheres in a 
light curve synthesis code}

\authorrunning{Orosz \& Hauschildt}

\maketitle

\begin{abstract}
We have written a light curve synthesis code that makes direct use
of model atmosphere specific intensities, in particular the {\sc NextGen}
model atmosphere grid for cool giants ($T_{\rm eff}\le 6800$~K and
$\log(g)\le 3.5$, Hauschildt et al.\ 
\cite{hauschildt99}).  We point out that these
models (computed using spherical geometry) predict a limb darkening
behaviour that deviates significantly from a simple linear or 
two-parameter
law (there is less intensity at the limb of the star). 
The presence of a significantly nonlinear limb darkening law has two
main consequences.  First, the ellipsoidal light curve computed for a
tidally distorted giant using the {\sc NextGen} intensities is in
general different from the light curve computed using the same
geometry but with the black body approximation 
and a one- or two-parameter
limb darkening law.  
In most cases the light curves computed with the
{\sc NextGen} intensities have deeper minima than their black body 
counterparts.
Thus the light curve solutions
for binaries with a giant component obtained with models
with near linear limb darkening (either black body or 
plane-parallel model atmosphere intensities)
are biased.  
Observations over a wide wavelength range
(i.e.\ both the optical and infrared) are particularly useful in
discriminating between models with nearly linear limb darkening and
the {\sc NextGen} models.
Second, we show that rotational broadening kernels for
Roche lobe filling (or nearly filling) giants can be significantly different
from analytic kernels due to a combination of the nonspherical
shape of the star and the radical departure from a simple limb darkening
law.  As a result, geometrical information inferred from $V_{\rm rot}
\sin i$ measurements of cool giants in binary systems are likewise biased.
\end{abstract}

\section{Introduction}
The  study of close binary stars
is of interest for several reasons.
For example, the understanding the structure and
evolution of stars is a basic goal of stellar astronomy, and is
required in most other branches of astronomy.  Critical tests of
evolution theory (i.e.\ predicting the radius and luminosity of a star
as a function of its mass and age) for stars other than the Sun are
practical only for a small set of eclipsing binary stars (see, e.g.,
Pols et al.\ \cite{pols97}; 
Schr\"oder et al.\ \cite{schroder97};
Lacy et al.\ \cite{lacy00}).  In addition, our
knowledge of the masses of stellar black holes (and many neutron stars
and white dwarfs as well) depends on our ability to interpret
ellipsoidal and usually non-eclipsing light curves (e.g.\ 
Avni \& Bahcall \cite{avni75}; 
Avni \cite{avni78}; 
McClintock \& Remillard \cite{mcclintock90}; 
Haswell et al.\ \cite{haswell93}; 
Shahbaz et al.\ \cite{shahbaz94}; 
Sanwal et al.\ \cite{sanwal96}; 
Orosz \& Bailyn \cite{orosz97}).  
Finally, considerable observational effort has been put forth
recently to use detached eclipsing binaries as extragalactic distance
indicators 
(Mochejska et al.\ \cite{mochejska98};
Ribas et al.\ \cite{ribas00}).  Since close binary stars are of such
importance, it is crucial that we have the ability to construct accurate
synthetic light curves for a variety of close binaries.

The light curve expected from a particular close binary depends on the
system geometry (i.e.\ the figures of the stars, their relative sizes,
separation, viewing angle, etc.) and on the radiative properties of
the stars, which are
set mainly by their effective temperatures, surface gravities 
and chemical compositions.  The equations describing the basic system
geometry for a close binary are
reasonably simple and have been known for a long time 
(e.g.\ see the text by 
Kopal \cite{kopal59}).  In practice, however,
the direct computation of light curves requires a computer, and codes
to compute light curves have been in use since the late 1960s (e.g.\
Lucy \cite{lucy67};
Hill \& Hutchings \cite{hill70}; 
Wilson \& Devinney \cite{wilson71}; 
Mochnacki \& Doughty \cite{mochnacki72}; 
Avni \& Bahcall \cite{avni75}; see also the review by Wilson
\cite{wil94}).
On the other hand, the detailed computation of stellar atmosphere models
which describe the specific intensity of radiation emitted by the
stellar surfaces is quite involved.  As a result, approximations are
frequently used in the computations of light curves.  The Planck
function is usually used to compute the normal monochromatic intensity
of a surface element with an effective temperature $T_{\rm eff}$:
\begin{equation}
I_0\propto[\exp(hc/k\lambda T_{\rm eff})-1]^{-1}
\end{equation}
where $\lambda$ is the effective wavelength of the observation and
$h$, $c$, and $k$ are the usual physical constants.  After
$I_0$ is found, the specific intensities for other emergent angles are
computed using a simple parameterisation (the ``limb darkening law''):
\begin{equation}
I(\mu) = I_0[1-x(1-\mu)]. 
\label{linear}
\end{equation}
The coefficient $x$ depends on the temperature, gravity, and chemical
composition of the star being modelled.  There are also many two-parameter
limb darkening laws including a ``quadratic'' law:
\begin{equation}
I(\mu) = I_0[1-x(1-\mu)-y(1-\mu)^2],
\label{quad}
\end{equation}
the ``square root'' law 
(D\'{\i}az-Cordov\'es \& Gim\'enez \cite{diaz92}):
\begin{equation}
I(\mu) = I_0[1-x(1-\mu)-y(1-\sqrt{\mu})],
\label{sqrt}
\end{equation}
or the ``logarithmic'' law 
(Klinglesmith \& Sobieski \cite{klinglesmith70}):
\begin{equation}
I(\mu) = I_0[1-x(1-\mu)-y\mu\ln\mu].
\label{log}
\end{equation}
There are many tabulations of the coefficients for these various laws
for a wide variety of temperatures, gravities, and chemical compositions
(e.g.\ Al-Naimiy \cite{alnaimiy78}; 
Wade \& Rucinski \cite{wade85}; 
Van Hamme \cite{vanhamme93};
Claret \cite{claret98}).

We discuss in this paper our technique for computing light curves
using model atmosphere specific intensities directly, thereby eliminating
the need for the black body approximation and the one- or two-parameter
limb darkening laws.  Although our emphasis here is on stars with low
effective temperatures and surface gravities
($T_{\rm eff}\le 6800$~K and $\log g\le 3.5$),
the technique is quite general.
In the next section we discuss some previous work in this area and
give details of our method.  We then discuss some of the basic results
and their implications, and give a summary of this work.  We have also
included
an appendix to the paper which gives some of the details of our
light curve code not directly related to the stellar atmospheres.

\section{Computational Technique}

\subsection{Previous Work}

The use of model atmosphere computations inside a light curve synthesis
code is by no means new.  The widely used Wilson-Devinney 
(\cite{wilson71}; hereafter
W-D) code has routine which applies a correction to the normal 
(black body) intensity:
\begin{equation}
I_0^{\rm corr}=I_0 r(T_{\rm eff}, \log g, \lambda),
\end{equation}
where $r(T_{\rm eff},\log g, \lambda)$ is the ratio of  
the filter-integrated stellar atmosphere
model characterised by $T_{\rm eff}$ and $\log g$ to the
filter-integrated
blackbody intensity.
The specific intensities for other angles
are then computed from $I_0^{\rm corr}$ using the limb darkening law.
The correction routine supplied with the W-D code is based on the
Carbon \& Gingerich (\cite{carbon69}) 
models.  R. E. Wilson (priv.\ comm.) informs
us he is in the process of updating the correction routine.  
Milone et al.\ (\cite{milone92}) 
have independently
written a correction routine for the W-D code based on 
the Kurucz (\cite{kurucz79}) models.  
Linnell \& Hubeny (\cite{linnell94}, \cite{linnell96})
have written a series of codes to compute synthetic spectra and light
curves for binary stars, including ones with disks.  They use Hubeny's
general spectrum synthesis code {\sc SYNSPEC} 
(Hubeny et al.\ \cite{hubeny94})
to generate the model spectra, using as input Kurucz (\cite{kurucz79}) 
atmosphere models for cool stars
($T_{\rm eff}\le 10000$~K) and {\sc TLUSTY} (Hubeny \cite{hubeny88})
and {\sc TLUSDISK}
(Hubeny \cite{hubeny91}) models
for hotter stars and disks, respectively.
They have applied
their model with some success to 
\object{$\beta$ Lyrae} 
(Linnell et al.\ \cite{linnell98a}) and 
\object{MR Cygni}
(Linnell et al.\ \cite{linnell98b}).  
Tjemkes et al.\ (\cite{tjemkes86}) have used 
Kurucz (\cite{kurucz79}) models
in their light curve synthesis code by tabulating filter-integrated
normal intensities for a grid of effective temperatures and surface gravities.
Specific intensities for other emergent angles are computed from
a limb darkening law tabulated from Kurucz models.  This code has been
applied successfully to X-ray binaries such as \object{LMC X-4}
(Heemskerk \& van Paradijs \cite{heemskerk89}) and \object{GRO J1655-40}
(van der Hooft et al.\ \cite{vanderhooft98}) 
and to the millisecond pulsar \object{PSR 1957+20}
(Callanan et al.\ \cite{callanan95}).
Shahbaz et al.\ (\cite{shahbaz94}) 
mention of the use of Kurucz (\cite{kurucz79}) fluxes in a code 
used at Oxford, but this paper does not specifically describe how
the model atmosphere fluxes are incorporated into the light curve
synthesis code.  Similarly, Sanwal et al.\ (\cite{sanwal96}) mention the use
of 
Bell \& Gustafsson (\cite{bell89}) model atmosphere fluxes
in the light-curve synthesis code developed
at the University of Texas at Austin 
(Zhang et al.\ \cite{zhang86};
Haswell et al.\  \cite{haswell93}) without giving specific details.

\subsection{Current Work}

We use a technique suggested to us by Marten van Kerkwijk to incorporate
model atmosphere intensities into our light curve synthesis code.  
In general, a detailed model atmosphere computation yields the 
specific intensity $I(\lambda,\mu)$ at a given wavelength $\lambda$
and 
emergent angle $\mu=\cos\theta$, where $\mu=1$ at the centre of
the apparent stellar disk and $\mu=0$ at the limb.  The total disk-integrated
intensity $I(\lambda)$
observed at the wavelength $\lambda$ is then
\begin{equation}
I(\lambda)= \int^1_0I(\lambda,\mu)\mu d\mu.
\end{equation}
Normally, the light curves of a binary star are observed in a broad bandpass.
The observed intensity $I_{\rm FILT}$
in a given bandpass is
\begin{eqnarray}
I_{\rm FILT} & = & \int^{+\infty}_{-\infty}I(\lambda)W_{\rm FILT}
(\lambda)
d\lambda  \cr
& = &
\int^{+\infty}_{-\infty}\left[\int^1_0I(\lambda,\mu)\mu d\mu\right]
W_{\rm FILT}(\lambda)
d\lambda,
\label{eq1}
\end{eqnarray}
where $W_{\rm FILT}$ is the wavelength-dependent transmission of the
filter bandpass in question.  We can reverse the order of the integrals
in Eq.\ (\ref{eq1}) to give
\begin{eqnarray}
I_{\rm FILT} & = & 
\int^1_0\left[\int^{+\infty}_{-\infty}I(\lambda,\mu)
W_{\rm FILT}(\lambda)
d\lambda
\right]\mu d\mu
  \cr
& = &
\int^1_0I_{\rm FILT}(\mu)\mu d\mu.
\label{eq2}
\end{eqnarray}
The quantity in square brackets in Eq.\ (\ref{eq2}), namely
\begin{equation}
I_{\rm FILT}(\mu)=\int^{+\infty}_{-\infty}I(\lambda,\mu)
W_{\rm FILT}(\lambda)d\lambda
\end{equation}
is independent of any geometry (for non-irradiated atmospheres), 
and as such can be computed in advance
of the light curve synthesis computations.

In our current implementation we compute
for each model characterised by a temperature $T_{\rm eff}$
and gravity $\log g$
a table of eight filter-integrated intensities for each specific angle
$\mu$.  The eight filters currently are the Johnson $UBVRIJHK$ system
where we have used the optical
filter response curves given in Bessell
(\cite{bessell90}) and the infrared filter response curves given
in Bessell \& Brett (\cite{bessell88}).  
Using all of the models we then generate
a table of the form

\begin{tabular}{l}
~              \\
{\tt T1~~~~g1} \\
{\tt Nmu             } \\
{\tt mu1~~I(1,1)~I(1,2)~...~I(1,7)~I(1,8)}~\\
{\tt mu2~~I(2,1)~I(2,2)~...~I(2,7)~I(2,8)}~\\
{\tt ...} \\
{\tt muN~~I(N,1)~I(N,2)~...~I(N,7)~I(N,8)}~\\
{\tt T1~~~~g2} \\
{\tt Nmu             } \\
{\tt mu1~~I(1,1)~I(1,2)~...~I(1,7)~I(1,8)}~\\
{\tt mu2~~I(2,1)~I(2,2)~...~I(2,7)~I(2,8)}~\\
{\tt ...    } \\
{\tt muN~~I(N,1)~I(N,2)~...~I(N,7)~I(N,8)}~\\
{\tt T2~~~~g1} \\
{\tt ...    } \\
~             \\
\end{tabular}

\noindent The first
two numbers are the temperature and gravity of the first model.
The next line gives the number of specific intensities that follow.
The next $N_{\mu}$ lines give the value of $\mu$ followed by
the eight filter-integrated intensities.  
The format is then repeated for the additional models.
The table is sorted in order
of increasing temperature, and for each temperature, the table is
sorted in order of increasing gravity.  

During the course of computing
a synthetic light curve the specific intensity of each surface element
must be specified.  These values 
of $I(T_{\rm inp},\log g_{\rm inp},
\mu_{\rm inp})$ are interpolated from the table using a simple
linear interpolation procedure.  First,
we locate the 4 nearest models ($T_{\rm up},\log g_{\rm up}$),
($T_{\rm up},\log g_{\rm low}$), ($T_{\rm low},\log g_{\rm up}$),
and ($T_{\rm low},\log g_{\rm low}$), where
$T_{\rm low}\le T_{\rm inp}\le T_{\rm up}$ and
$\log g_{\rm low}\le \log g_{\rm inp}\le \log g_{\rm up}$.  Next, within
each of the 4 nearest models we find the filter-integrated intensities
appropriate for  $\mu_{\rm inp}$, using linear interpolation.
Finally, the desired values of \\
$I(T_{\rm inp},\log g_{\rm inp},
\mu_{\rm inp})$ are found by linear interpolation first in the $\log g$
direction and then the $T_{\rm eff}$ direction.  Returning the
filter-integrated intensities for eight different filters at once 
is advantageous when modelling observations in several bandpasses since
one has to search the intensity table only once per surface element
per phase.

\begin{figure}
\centering
\centerline{\epsfxsize=8.2cm
\epsfbox{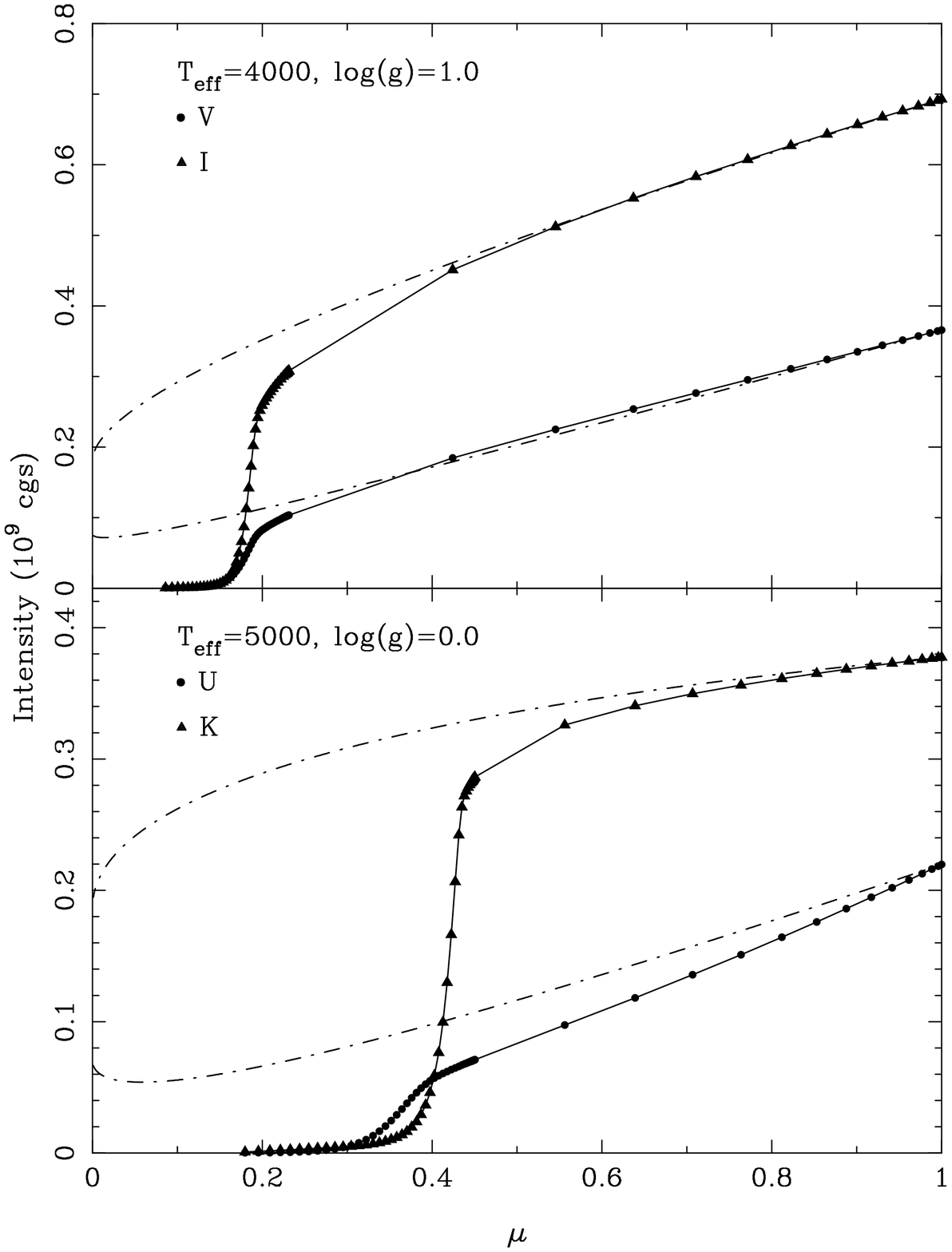}}
\caption{Top:  the filter-integrated specific intensities
$I_{\rm FILT}(\mu)$ for Johnson $V$ and $I$
for a model with $T_{\rm eff}=4000$~K and
$\log g=1$.  The dashed-dotted lines show the limb darkening
behaviour predicted by the ``square root'' law (Eq.\ (\ref{sqrt})),
where we have used coefficients (computed from Kurucz models) given
in Van Hamme (\protect\cite{vanhamme93}).  Bottom:
the filter-integrated specific intensities
$I_{\rm FILT}(\mu)$ for Johnson $U$ and $K$
for a model with $T_{\rm eff}=5000$~K and
$\log g=0$, and the corresponding square root limb darkening laws.
}
\label{fig1}
\end{figure}

\begin{figure}
\centering
\centerline{\epsfxsize=8.2cm
\epsfbox{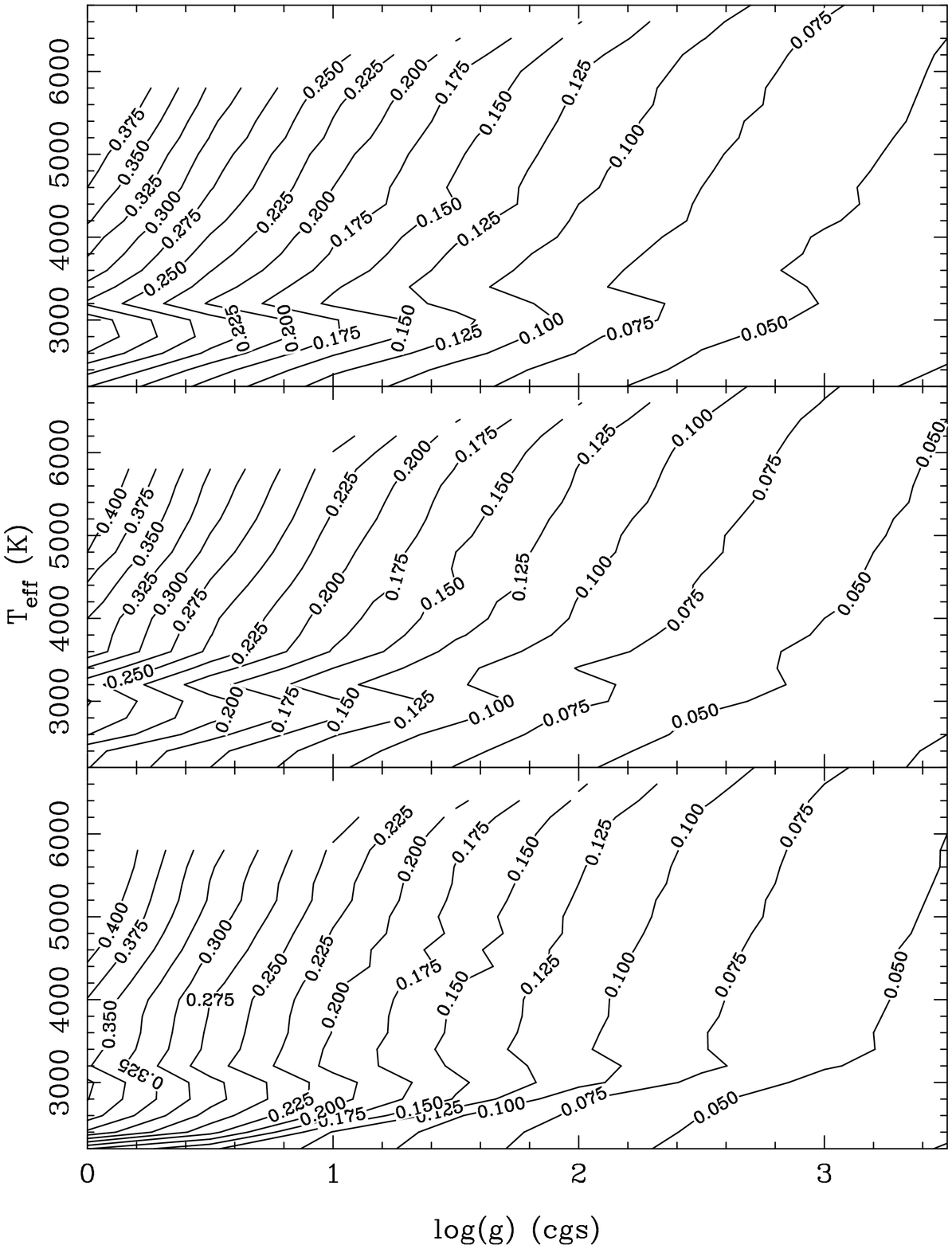}}
\caption{Contour plots of the quantity $\mu_{\rm cut}$ for the
{\sc NextGen} models in
the $B$ (top),
$I$ (middle), and $K$ (bottom) bands.  There are no static models
for the highest temperatures and lowest gravities, hence there are gaps
in the upper left parts of the contour maps.
}
\label{fig3}
\end{figure}

\begin{figure}
\centering
\centerline{\epsfxsize=8.2cm
\epsfbox{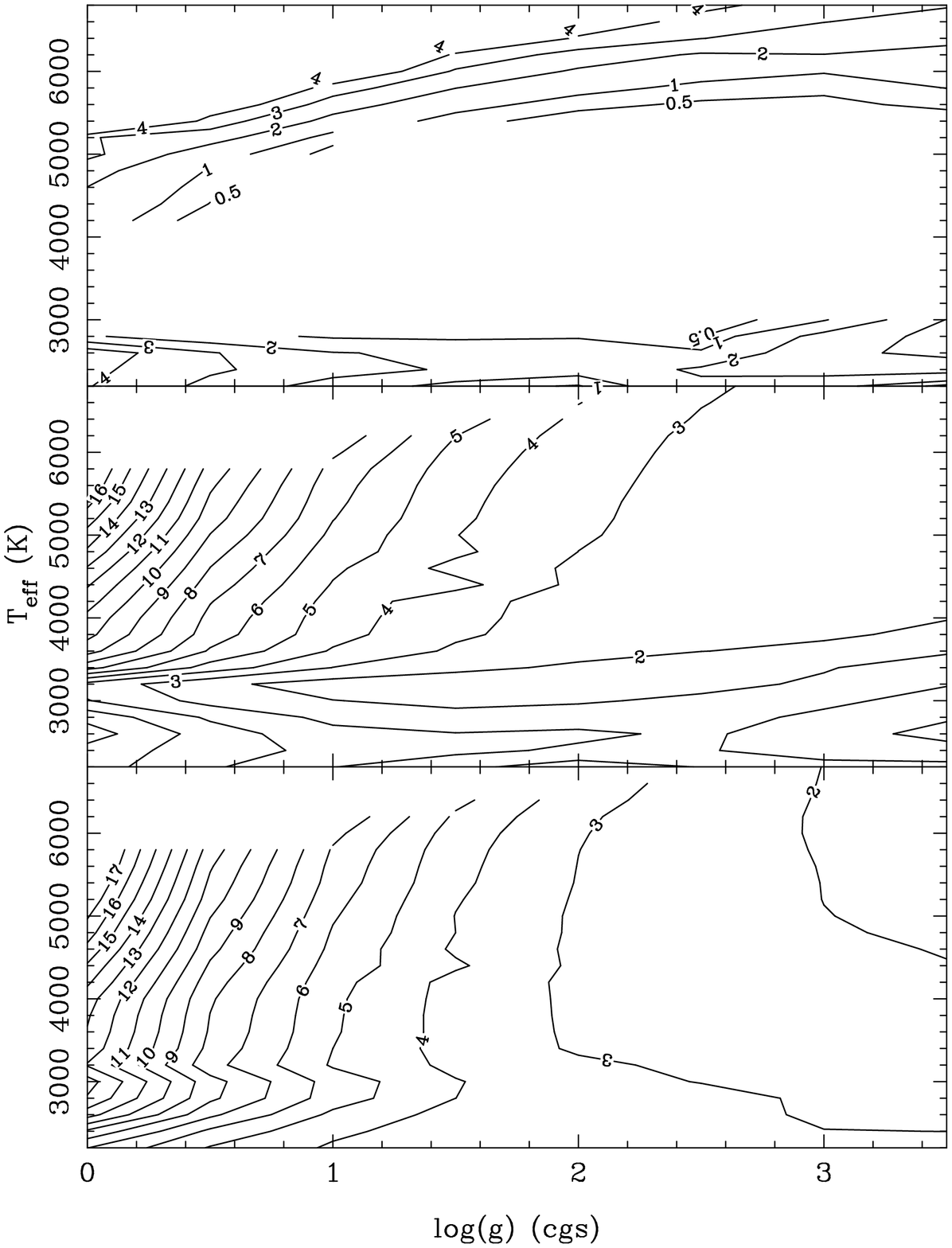}}
\caption{Contour plots of the ``lost light,'' expressed as
a per cent, of the 
{\sc NextGen} models in
the $B$ (top),
$I$ (middle), and $K$ (bottom) bands.  
There is a large area of the parameter space for the
$B$ filter (shown as the large gap in the contours)
where there is no lost light as defined in the text.
}
\label{figlost}
\end{figure}

As we noted above, the technique we just outlined is quite general.
We now turn our attention to the recent grid of
{\sc NextGen} models
for cool giants computed by PHH using the parallelised version of
his general purpose code
{\sc PHOENIX} (Hauschildt et al.\ \cite{hauschildt97}).
The computational techniques and the input physics used to
compute the {\sc NextGen} models for cool giants have been
discussed elsewhere (Hauschildt et al.\ \cite{hauschildt99}) and will not
be repeated here.   These spherical models require an input mass
to compute the sphericity factor,
and we use $M=5\,M_{\odot}$ for all of the models.  We did examine
a few models with masses of $2.5\,M_{\odot}$ and 
$7.5\,M_{\odot}$ and found no discernible differences
compared to the $M=5\,M_{\odot}$ models. 
For the moment we restrict ourselves to the models with solar
metallicity.  For each model the
specific intensities are computed for 64 different angles over a wavelength
range of 3000-24,998~\AA\ in 2~\AA\ steps.  The distribution of the
emergent angles $\mu$ is chosen by the {\sc PHOENIX}
code based on the structure of the atmosphere under consideration.  Hence, 
in general, the 64 values of $\mu$ are different from model to model.

Fig.\ \ref{fig1} shows the quantity $I_{\rm FILT}(\mu)$ for two different
{\sc NextGen}
models and filter combinations.  The top panel shows the integrated
intensities in the Johnson $V$ and $I$ filters for the
model with $T_{\rm eff}=4000$~K and $\log g=1.0$ and the
bottom panel shows  
the integrated
intensities in the Johnson $U$ and $K$ filters for the
model with $T_{\rm eff}=5000$~K and $\log g=0.0$.
For comparison, we also show
the square root limb darkening laws (Eq.\ (\ref{sqrt})) computed using
coefficients taken from the tabulation of 
Van Hamme (\cite{vanhamme93}).  These four
curves essentially show the limb darkening behaviour predicted by the
Kurucz models
to within a few per cent, although 
is important to note that the intensities for $\mu=1.0$ are 
different between the Kurucz models and the {\sc NextGen} models (see
Hauschildt et al.\ (\cite{hauschildt99}) for further discussion of this point).
This figure is quite striking.
It shows the radically different limb darkening behaviour predicted by
the spherical {\sc NextGen} models and the plane-parallel Kurucz models.
The {\sc NextGen} model predicts 
a sharp decrease in the intensity at relatively large $\mu$ values
(as large as $\approx 0.4$),
whereas the limb darkening parameterisations predict substantial 
intensity all the way to $\mu=0$.
This sudden decrease in the
intensity is a consequence of the spherical geometry:  for sufficiently
low gravities ($\log g\la 3.5$ for most temperatures) there is much
less material near the limbs and hence much less radiation.  The radiation
that would have come out near the limb in the redder bandpasses comes
out at much shorter wavelengths (see 
Hauschildt et al.\ \cite{hauschildt99}).

The value of $\mu$ where the sudden fall-off in the intensities
occurs depends on the effective temperature and gravity of the stellar
model, and on the bandpass.  We define the slope 
of the intensities at  $\mu(i)$ to be
\begin{equation}
s(i)={I(i+1)-I(i-1)\over \mu(i+1)-\mu(i-1)}
\end{equation}
The location $\mu_{\rm cut}$
of the edge
is defined to be the point where $s(i)$ is the largest.
Fig.\ \ref{fig3} shows contour plots of $\mu_{\rm cut}$
for three different bands (Johnson $B$, $I$, and $K$).  
In general, for a fixed effective temperature, 
$\mu_{\rm cut}$ gets larger as the gravity of the model
decreases.
Also, for a fixed  $\log g$, the hotter models tend to have a slightly
larger values of $\mu_{\rm cut}$.
There is 
some irregular behaviour
seen in the contours of $\mu_{\rm cut}$ at temperatures
between about 3000~K and 4000~K.  
The reason for this seems to be that 
for models in this temperature range,
the 
fall-off in the intensity is generally less sudden, especially
in the bluer filters.
Thus $\mu_{\rm cut}$ is less clearly defined in these cases.

\begin{figure*}
\vspace{9.5cm}
\includegraphics{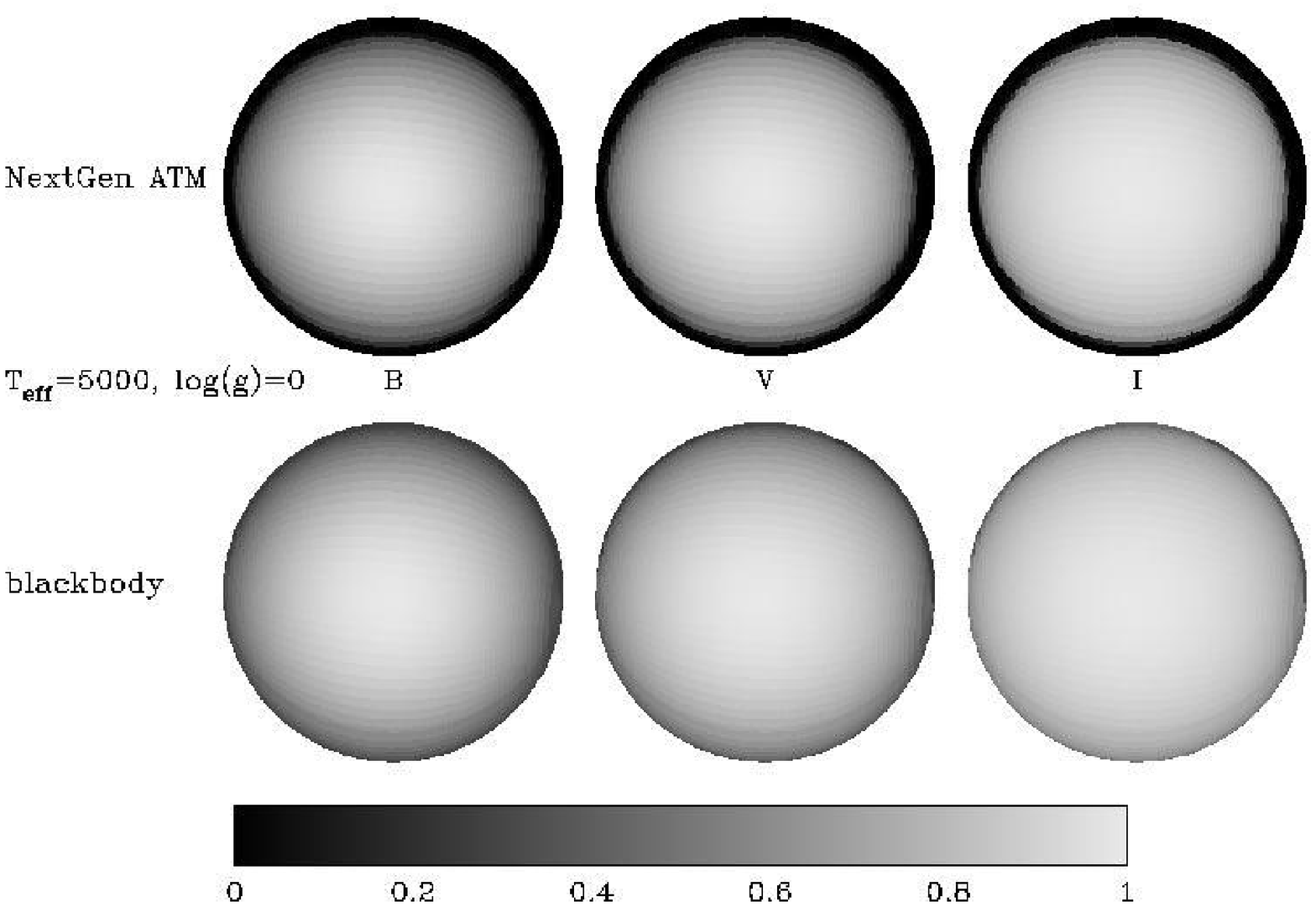}
\caption{Intensity maps for a spherical star with
$T_{\rm eff}=5000$~K and $\log g=0.0$.  The top images show intensities
computed from the {\sc NextGen} models in the Johnson $B$, $V$, and $I$
bandpasses, while the bottom images show monochromatic
black body
intensities with the square root limb darkening law (Eq.\
(\ref{sqrt})) at the effective wavelengths of 
the same three filters.  In each case the intensities
have been normalised to 1.0 at the disk centre.
}
\label{fig2}
\end{figure*}

It is quite obvious that the quantity $I_{\rm FILT}$ (Eq.\
(\ref{eq2})) evaluated for the {\sc NextGen} intensities
will be less than the corresponding quantity evaluated
for a one- or two-parameter limb darkening law.  
For each model and
filter,
we computed the ``lost light'' in the following way.  We first  
``linearised'' our
table of {\sc NextGen} intensities:  We fitted
a line to the  intensities
for the ten specific angles nearest to $\mu=1.0$.  Then the specific
intensities for the remaining 54 angles were replaced by the intensities
determined from the extrapolation of the fitted line.  In most cases,
the extrapolated intensities at the limb were above zero.  There were
a few cases
such as the $T_{\rm eff}=5000$, $\log g=0.0$ $U$-band model
shown in Fig.\ \ref{fig1} where the extrapolated intensities
at the limb would have been negative.  In such cases we replaced all negative
intensities with zero.
Eq.\ (\ref{eq2}) was evaluated for the linearised table and the
regular table, and we define the lost light as $D(T_{\rm eff},
\log g)$ $=$ $100\times(I_{\rm FILT,linear}(T_{\rm eff},
\log g)-I_{\rm FILT,regular}(T_{\rm eff},
\log g))$ $/$ $I_{\rm FILT,linear}(T_{\rm eff},
\log g)$.  Fig.\ \ref{figlost} shows contour plots of 
\\ $D(T_{\rm eff},
\log g)$ for the $B$ (top), $I$ (middle), and $K$ (bottom) filters.
By our definition the lost light in many of the $B$-band models is
zero.  Otherwise, for the $B$ band,
the lost light is typically a few per cent and
about 3 or 4 per cent for the hottest models.
On the other hand, the lost light in the $I$ and $K$ bands can be quite
substantial, up to 17 per cent for the models with lowest gravities and
the hottest temperatures.

We can easily make an image of the disk a model star as it would appear
in the sky.
The intensity maps displayed in Fig.\ \ref{fig2} compare the
{\sc NextGen} intensities for $T_{\rm eff}=5000$~K and $\log g=0.0$
to monochromatic black body intensities (limb darkened 
using the square root law (Eq.\ (\ref{sqrt})) for three different
filters
(Johnson $B$, $V$, and $I$).  
The star with the {\sc NextGen} intensities appears smaller
on the sky.  The difference in the limb darkening closer to the disk
centre is also apparent from the figure.

Finally, for comparison purposes, we generated an intensity table from
Kurucz models.  We retrieved the file ``ip00k2.pck19'' from Kurucz's
public World Wide Web pages\footnote{http://cfaku5.harvard.edu/}.  This
file contains the specific intensities for 17 different angles
for a grid of solar abundance models.  The coolest models have 
$T_{\rm eff}=3500$~K.  All of these models are plane-parallel and LTE.

\section{General Results}

We have written a new light curve synthesis code which is based on
the work of Avni 
(Avni \& Bahcall \cite{avni75}; Avni \cite{avni78}).  This code,
named ELC, can model any semi-detached or detached binary with
a circular orbit (we will generalise the code to include  overcontact
binaries and binaries with eccentric orbits within the next year).
The second star can be surrounded by an accretion disk.  We will
defer giving specific details of the code to the Appendix and now
turn our attention to the results related to the use of the
{\sc NextGen} intensities.

\subsection{Light curve amplitude}

\begin{figure}
\centering
\centerline{\epsfxsize=8.2cm
\epsfbox{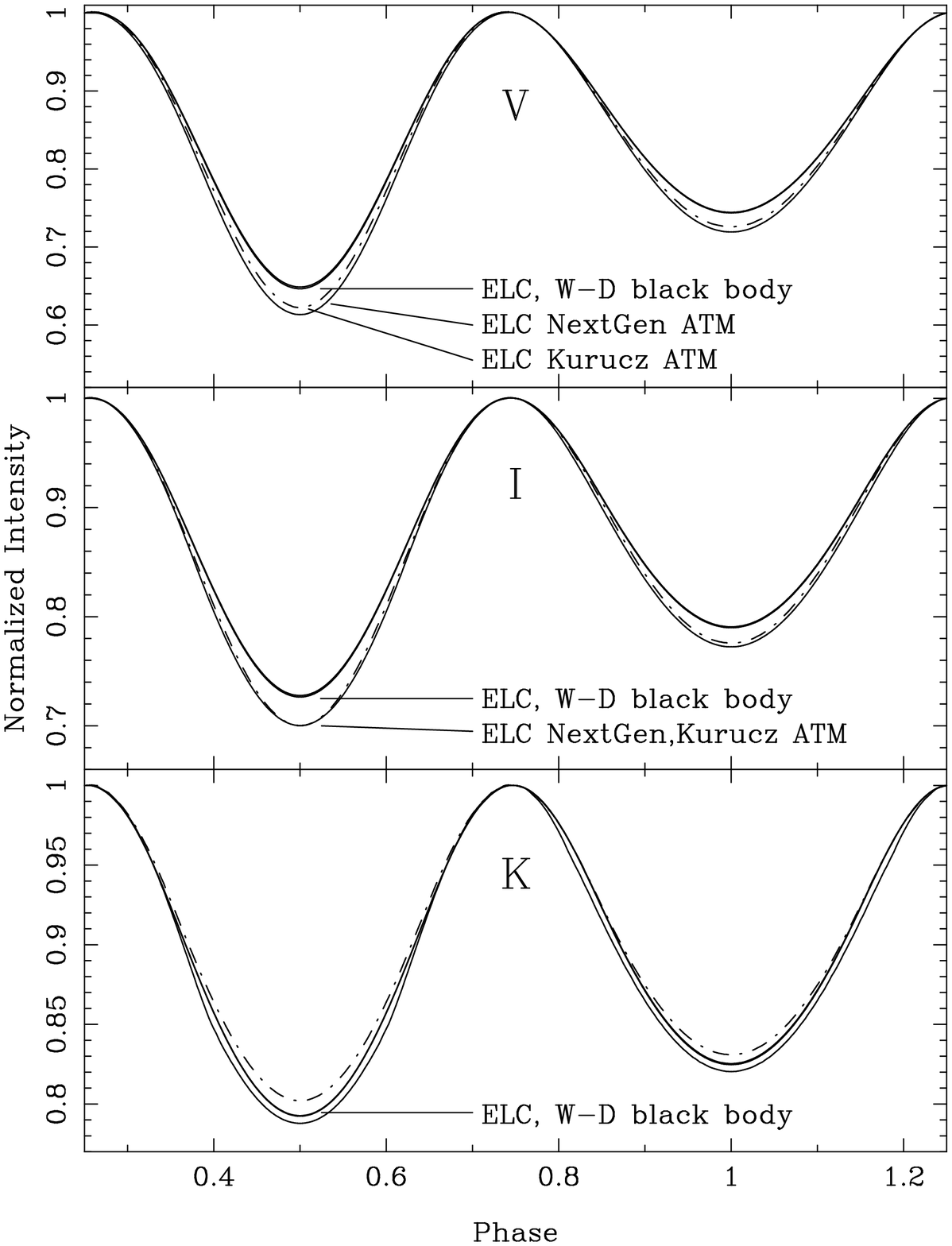}}
\caption{Top: $V$-band light curves for a single Roche lobe filling
star computed in four
ways.  The mean temperature of the star is 
$T_1=4000$~K, its surface
gravity is $\log g=1.0$, the binary mass ratio is $Q=5$,
and the inclination is $i=75$ degrees.  The $y$-axis
is an intensity scale.
The two curves with the largest intensities at phase 0.5
are the ELC and W-D
monochromatic curves computed using
black body intensities (the two curves are nearly identical).
The curve in the middle drawn with the dash-dotted
line is the light curve generated by ELC using
the Kurucz intensity table.  The curve with the lowest intensity
at phase 0.5 is the ELC curve computed using the {\sc NextGen}
intensities.
Middle: Same as the top, but for the $I$ filter.  In this case
the curves computed using the Kurucz and {\sc NextGen}
intensities are nearly the same.
Bottom:  Same as the top, but for the $K$ filter.
The ELC {\sc NextGen} curve has the deepest minimum at phase
0.5.
}
\label{fig4}
\end{figure}

For various test cases we computed light curves in four different ways:
using our ELC code in the black body mode, using the W-D code in the
black body mode (we have added a phase shift of 0.5 to the W-D models,
see the Appendix), 
using our ELC code with the Kurucz intensity table, and
using our ELC code with the {\sc NextGen} intensity table.
Fig.\ \ref{fig4} shows four
different normalised light curves (in $V$, $I$, and $K$)
for a single Roche lobe filling
star in synchronous rotation
with $T_{\rm eff}=4000$~K and $\log g=1.0$.  The mass ratio 
is 5, meaning the unseen second star is five times more massive
than the cool giant.  The inclination is 75 degrees.
For the black body models we used the square root limb darkening law with
coefficients from Van Hamme (\cite{vanhamme93}). 
The gravity darkening exponent was
$\beta=0.08$, appropriate for a star with a convective envelope. 
(We note that the black body light curves were computed
using a single limb darkening law for the entire star, whereas the light
curves computed using model atmosphere intensities ({\sc NextGen} or
Kurucz) will have location-specific limb darkening ``built in''.)
In all three cases, the two black body curves
are nearly identical (the largest deviations between the two are on the
order of 2 millimags).  
Typically the
{\sc NextGen} curves have deeper minima than the black body  curves.
Depending on the filter, the {\sc NextGen} light
curves are different from the
the light curves
computed with Kurucz intensities, although for the example shown in
Fig.\ \ref{fig4} the two $I$-band curves have only minor differences.

\begin{figure}
\centering
\centerline{\epsfxsize=8.2cm
\epsfbox{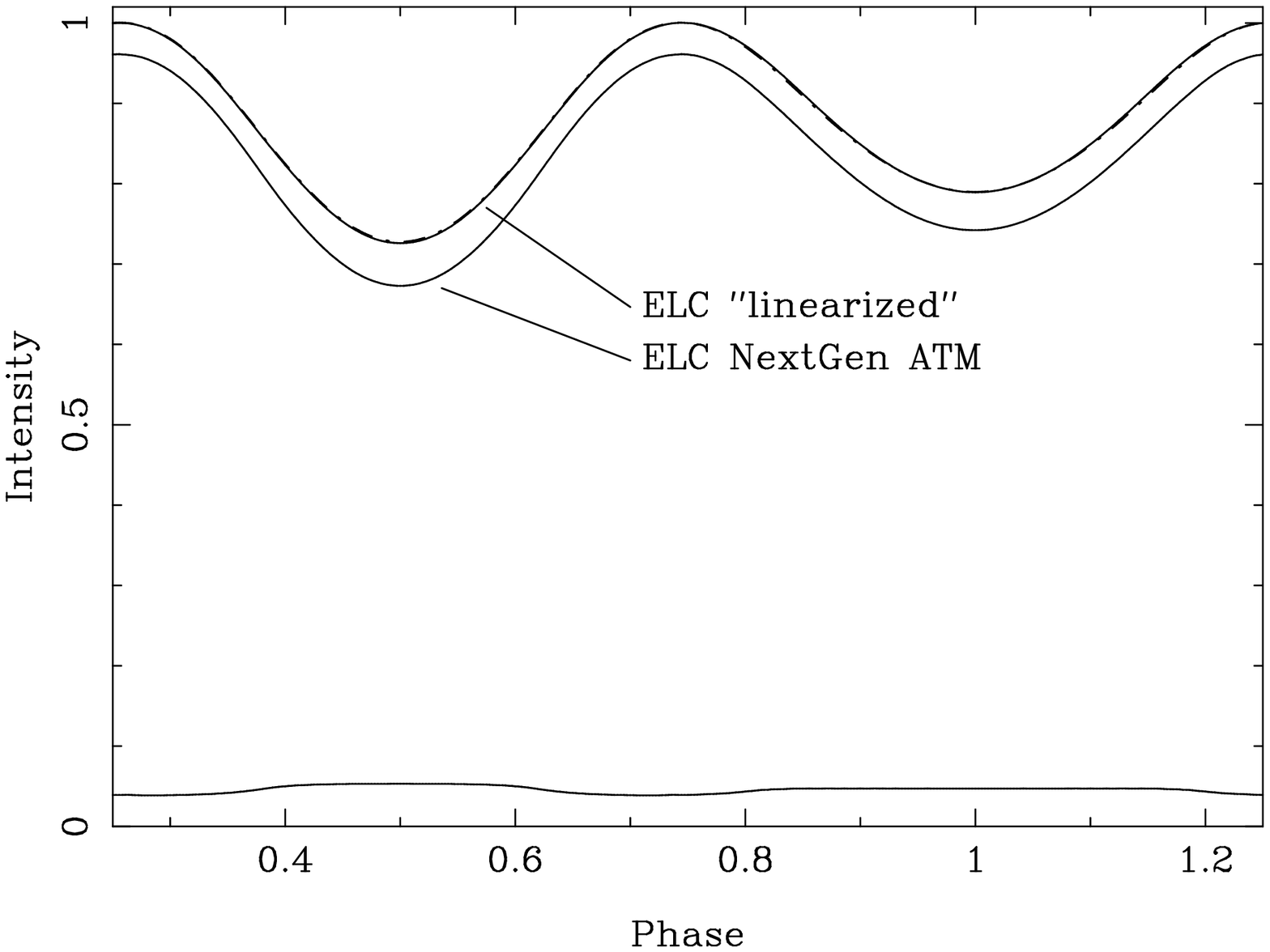}}
\caption{The top curve is the $I$-band light curve
for a single Roche lobe filling star with $T_{\rm eff}=4000$~K
and a mean gravity of $\log g=1.0$ constructed using the ``linearised''
intensity table.   As in Fig.\ \protect\ref{fig4}, $i=75^{\circ}$ and
$Q=5$.  The $I$-band black body light curve shown in Fig.\
\protect\ref{fig4} is overplotted with the dash-dotted line (there
is essentially no difference between the two curves).  The
middle curve is the light curve made using the regular
{\sc NextGen} intensity table, plotted on the same intensity scale.
The bottom curve is the difference between the linearised curve and the
regular curve. 
}
\label{fig5}
\end{figure}

In general, the light curves computed using the {\sc Next}{\sc -Gen} intensity
table are different from light curves computed using the same geometry
but with black body intensities.  This comes as no surprise, given the
potentially large differences in the limb darkening between the
{\sc Next}{\sc -Gen} 
models and the one- or two-parameter limb darkening laws.
It is usually the case (especially in the
$I$, $J$, $H$, and $K$ filters)
that the {\sc NextGen}
light curves have larger amplitudes (in normalised intensity or in magnitudes)
than the corresponding black body light 
curves.  Fig.\ \ref{fig5} shows why this is so.  The top curve
is the $I$-band light curve of the single star described
in the Fig.\ \ref{fig4} caption made with the linearised
intensity table.  Since the square root limb darkening law for this
model ($T_{\rm eff}=4000$, $\log g=1.0$) and bandpass closely
matches the actual {\sc NextGen} intensities near disk centre
(see Fig.\ \ref{fig1}), the corresponding black body light curve
is nearly identical to the ``linearised'' light curve.
The middle curve in Fig.\ \ref{fig5} is the light curve for the
same star made with the regular {\sc NextGen} intensity table,
plotted on the same intensity scale.  The average light level in
this light curve is about 5 per cent lower than the linearised
one, which
is roughly what we would expect from Fig.\ \ref{figlost}.  The bottom
curve is the difference between the two top light curves.  This lower
curve is basically the light lost near the limb, and it is essentially
constant with phase.  The two top curves have roughly the same amplitude
in these intensities units.  That is, $I_{\rm max}-I_{\rm min}$
is the same for both curves.  However, the {\em relative} amplitude
$(I_{\rm max}-I_{\rm min})/I_{\rm ave}$ for the regular {\sc 
NextGen} light curve
will be larger since its mean light level is lower.
Hence, the amplitude of this particular {\sc NextGen} light curve
(and most others) in {\em normalised intensity or in magnitudes}
is larger than the amplitude of the corresponding black body curve.

There are some situations such as the $U$ model shown in the
bottom of Fig.\ \ref{fig1} where the square root limb darkening law
does not match the slope of the {\sc NextGen} intensities near $\mu=1$.
For cases like this, the black body light curve can have a larger amplitude
than the {\sc NextGen} light curve.  

\begin{figure}
\centering
\centerline{\epsfxsize=8.2cm
\epsfbox{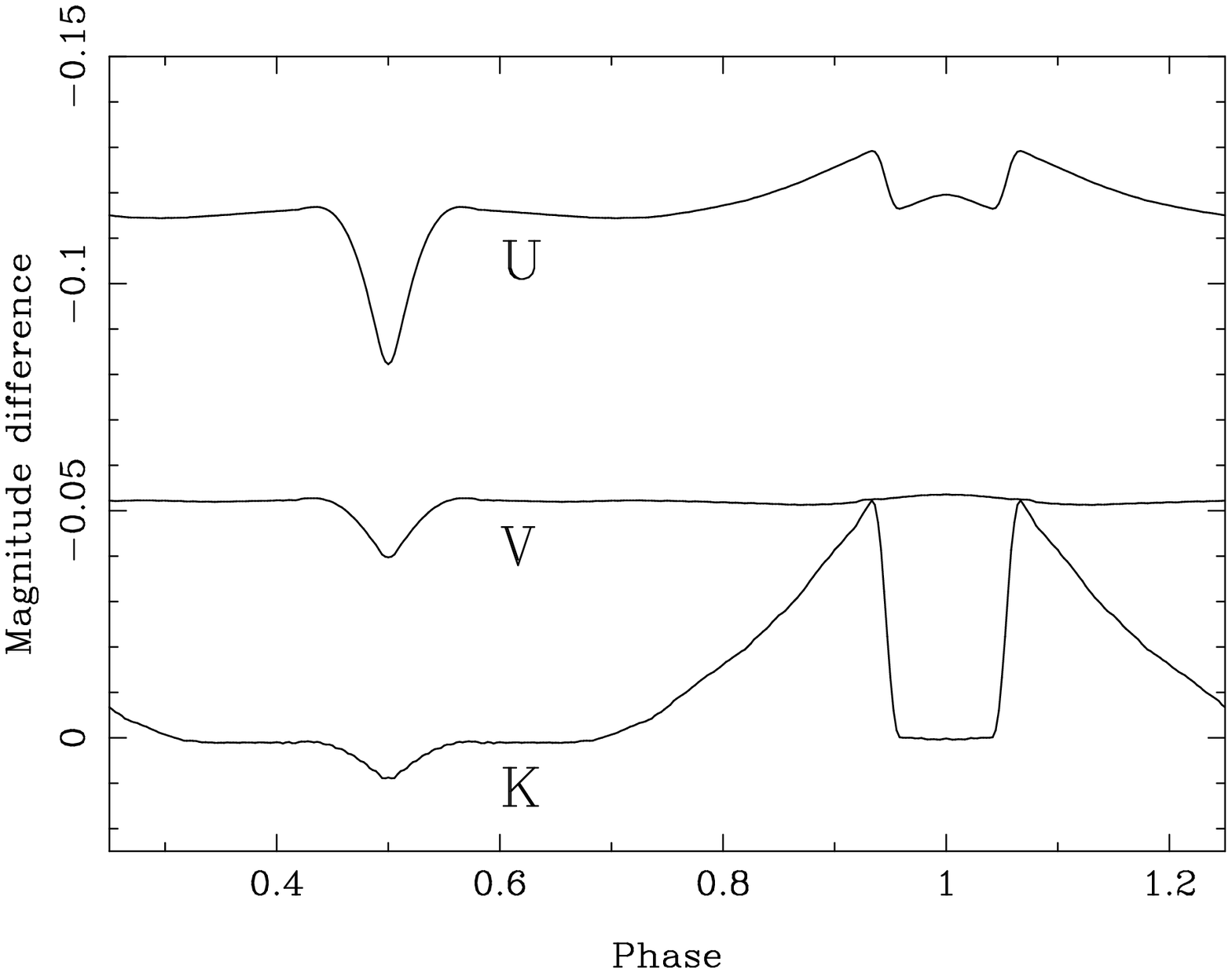}}
\caption{The difference curves {\em in magnitudes} between
the {\sc NextGen} light curves and the Kurucz light curves for
a binary with a geometry
similar to \protect\object{RZ  Scuti}
(i.e.\ a Roche lobe-filling ``cool'' star and a 
``hot'' mass-gaining star rotating near its critical limit), but
with altered temperatures.  
The adopted parameters are:
$i=85.65^{\circ}$,
$Q=0.216$,
$P=15.2$ days,
$a=61.89\,R_{\odot}$,
$T_{1}=6000$~K,
$T_{2}=5000$~K,
$f_1=0.66045$,
$f_2=1.0$,
$\Omega_1=4$,
$\Omega_2=1$,
$\beta_1=0.25$,
$\beta_2=0.08$,
and detailed reflection with albedos of 1.0 and 0.5 for the hotter and
cooler star, respectively (see the Appendix for a detailed discussion
of the free parameters of the model).  The resulting physical masses,
radii, and mean gravities of the two stars are 
$M_1=11.3\,M_{\odot}$, $R_1=14.4\,R_{\odot}$, $\log g_1=3.17$ and
$M_2=2.45\,M_{\odot}$, $R_2=16.2\,R_{\odot}$, $\log g_2=2.42$, 
respectively.
}
\label{rzscuti}
\end{figure}

Fig.\ \ref{fig4} nicely shows the difference in the light curves between
black body intensities and model atmosphere intensities (e.g.\ either
{\sc NextGen} or Kurucz).  
However, in the example shown, the 
{\sc NextGen} light curves and the Kurucz light curves had only
minor differences.  To illustrate how large the difference between
a {\sc NextGen} light curve and a Kurucz light curve can be we consider
a binary similar to \object{RZ Scuti}.  This semidetached binary consists of
a B3Ib star which rotates near its centrifugal limit and a
Roche lobe-filling A2? star (e.g.\ Olson \& Etzel \cite{oe94}).  
The mean gravities of the two stars are $\log g_1\approx 3.2$ and
$\log g_2\approx 2.4$, so sphericity effects should be important here.
As a result
of its rapid rotation near the critical limit
(see Van Hamme \& Wilson \cite{vw90} for a discussion of ``critical
rotation''), 
the  
B-star is significantly flattened.  Furthermore, the surface gravity
near its equator is relatively low, giving rise to a relatively large
range in temperatures owing to the increased gravity darkening.  
We currently cannot model the real \object{RZ Scuti} binary
using the {\sc NextGen} intensities because the mean temperature of
the B-star is well outside our current model grid.  Indeed, there 
are parts on the B-star where the temperature and gravity combination
fall outside the Kurucz grid, so we cannot model the real binary with
the Kurucz grid either.  For this example 
we therefore modified the temperatures of the
two stars and ``slowed down'' the mass gaining primary slightly
so that the
temperature and gravity combination of each point on each star is contained
in both the {\sc NextGen} and Kurucz grids.  The adopted model parameters
are summarised in the caption of Fig.\ \ref{rzscuti}.    
Fig.\ \ref{rzscuti} itself shows the
difference between the {\sc NextGen} light curve and the Kurucz
light curve {\em in magnitudes} for three filters.  There are large
differences between the two models, and the size of the difference
depends on the filter bandpass.
The difference
between  $K$-band
light curves is as large as 0.05 mag  near phase 0, when the cooler star
is eclipsed by the hotter star.  One can also note from Fig.\ \ref{rzscuti}
that the $U$ and $V$ difference curves have means which are less than zero.  
In other words the
binary is {\em bluer} in $U-K$ and $V-K$ when {\sc NextGen} intensities
are used compared to when Kurucz intensities are used.

\begin{figure}
\centering
\centerline{\epsfxsize=8.2cm
\epsfbox{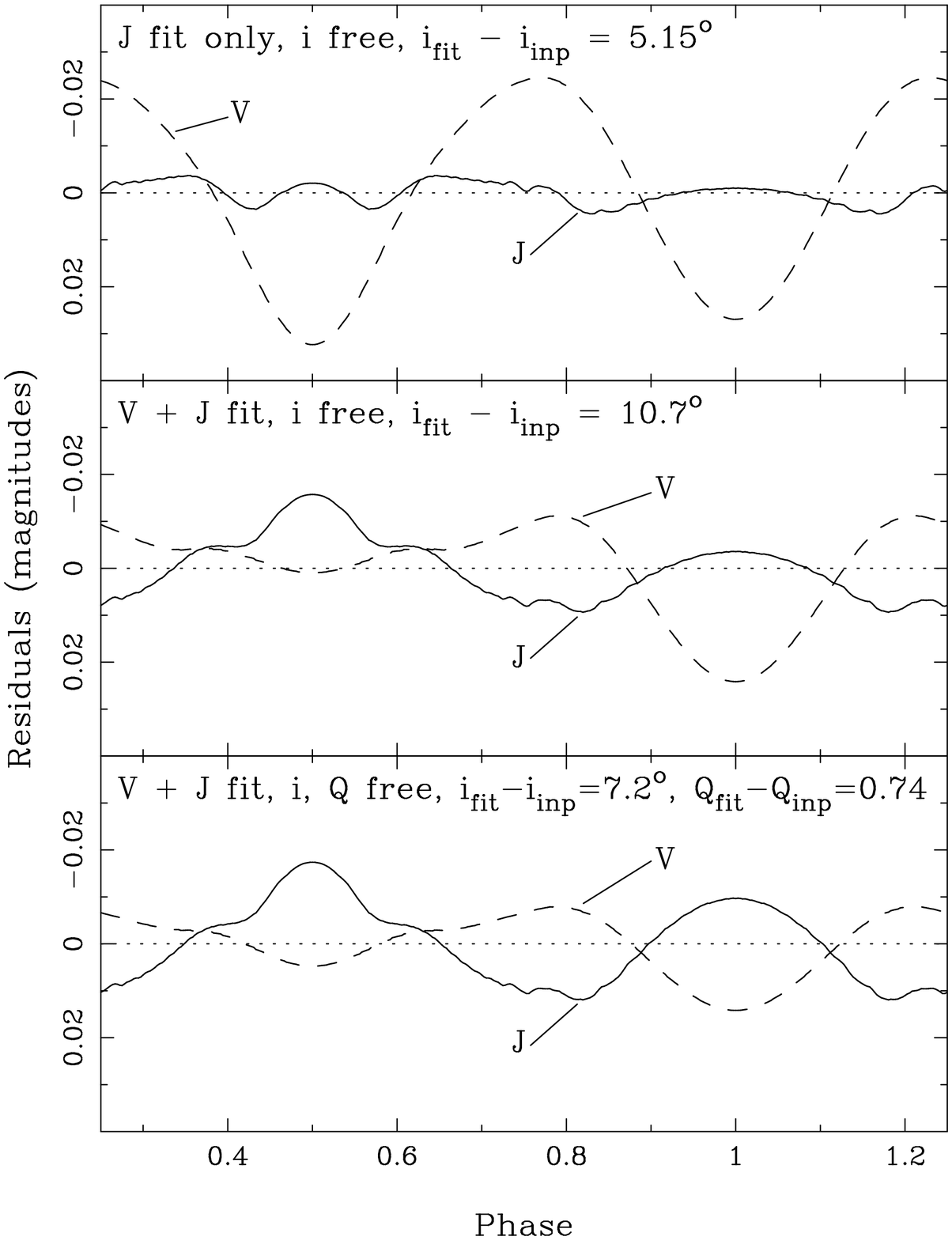}}
\caption{The results of fitting a {\sc NextGen} light curve
using black body intensities.  The input model parameters
are single Roche lobe-filling star,
$T_{\rm eff}=3560$~K, $Q=1.6666$, and $i=60^{\circ}$.
Top panel: the $O-C$ residuals (in magnitudes) of a fit to the
$J$-band light curve only ($Q$ fixed, $i$ free).  The fitted inclination
is $i=65.15^{\circ}$, and the amplitude of
the $V$-band light curve is nearly 0.03 magnitudes too small.
Middle panel: similar to top, but a fit to both $V$ and $J$.  The fitted
inclination is nearly $11^{\circ}$ too large, and the $V$ residuals
still have relatively large systematic deviations.
Bottom panel:  similar to the middle, but both the inclination and
mass ratio were free parameters.  The residuals are reasonably small,
but the fitted parameters are quite different from the input parameters.
}
\label{fitfig}
\end{figure}

To quantify how much the difference in the light curve amplitudes
between the {\sc NextGen} models and the black body models might matter
when fitting the light curves of a real binary star,
we consider a model binary similar to 
\object{T Coronae Borealis}.  
\object{T CrB} is an S-type symbiotic binary where an M4 giant orbits
an optically faint hot companion with a period of about 227.6
days 
(Kraft \cite{kraft58}; 
Kenyon \& Garcia \cite{kenyon86}; 
Fekel et al.\ \cite{fekel00}).  There are no
eclipses observed in the UV 
(Selvelli et al.\ \cite{selvelli92}).
Nevertheless, the large amplitude of the ellipsoidal light curves suggests
the M giant fills its Roche lobe and is viewed at a large inclination
(Bailey \cite{bailey75};  
Lines et al.\ \cite{lines88}; 
Yudin \& Munari \cite{yudin93}).
Since the density of a Roche lobe filling star is a function only
of the orbital period to a good approximation 
(Faulkner et al.\ \cite{faulkner72}; 
Eggleton \cite{eggleton83}), the surface gravity of the M giant
is only a weak function
of its assumed mass.  In the case of 
\object{T CrB}, $\log g\approx 0.7$, so we
expect the sphericity effects to be important here. 
The mass ratio of the binary is not well known.  
Kraft (\cite{kraft58}) measured
hydrogen emission line radial velocities on seven plates and found
$M_{\rm giant}/M_{\rm comp}\approx 1.4$.  However, this mass ratio implies
a rotational velocity of $V_{\rm rot}\sin i\approx 23$ km s$^{-1}$,
which
is much larger than the upper limit of $\approx 10$ km s$^{-1}$
measured by Kenyon \& Garcia (\cite{kenyon86}).  We discuss 
below some
of the potential problems associated with measuring the rotational
velocity of a tidally distorted cool giant, and in view of this discussion
the rotational velocity upper limit of Kenyon \& Garcia (\cite{kenyon86}) 
should be treated with caution.

The $J$ and $K$ band light curves of T CrB collected between
1987 August and 1995 June are stable and representative of the
true ellipsoidal modulation 
(Yudin \& Munari \cite{yudin93}; 
Shahbaz et al.\ \cite{shahbaz97}).  
The full amplitude of the $J$ light curve is about 0.2 magnitudes.
On the other hand, the $V$ light curve shows additional sources of
variability not associated with the underlying ellipsoidal
modulation (e.g.\ Lines et al.\ \cite{lines88}).  
Nevertheless, between about 1989 and
the beginning of 1996, the $V$ light curve seemed to be reasonably stable.
Hric et al.\ (\cite{hric98}) refer to this period as the ``quiet stage''.  
The full amplitude of the $V$ light curve during this quiet stage is
about 0.4 magnitudes.  For this discussion we will assume this light
curve represents the true ellipsoidal light curve.  
Both Shahbaz et al.\ (\cite{shahbaz97}) and 
Belczy\'nski \& Mikolajewska (\cite{belczynski98})
had difficulty fitting the  optical ($V$ or $I$) light curves
simultaneously with the infrared ($J$ or $K$) light curves.   
If the amplitude of the $J$ band light curve was matched, then
the model $V$ and $I$ light curves (computed using black body intensities)
had amplitudes that were about a factor of two too small.  

A thorough analysis of existing 
\object{T CrB}  data is beyond the scope of the
present paper and will be deferred to a future paper.  For now
it will suffice to discuss simulated T CrB-like light curves.  For these
light curves we will use the 
\object{T CrB} system parameters adopted by Belczy\'nski \& Mikolajewska 
(\cite{belczynski98}),
namely $M_{\rm giant}/M_{\rm comp}=0.6$, which in our notation is
$Q=1/0.6=1.6666$, 
$i=60^{\circ}$, 
$T_{\rm eff}=3560$~K, and a gravity darkening exponent of
$\beta=0.08$, appropriate for a star with a convective outer envelope.
We also used the orbital period and $K$-velocity of the M giant given
in Kenyon \& Garcia (\cite{kenyon86}).  
Using the {\sc NextGen} intensity table
we computed light curves for the Johnson $V$ and $J$ filters.  The
model light curves were  converted to magnitudes for compatibility with
our optimising  routines.  We used ELC in its black body mode to fit
various combinations of the simulated light curves, and the results are
shown in
Fig.\ \ref{fitfig}.  The top panel shows the results of fitting the $J$
light curve only, using the inclination as the sole free parameter.
The fitted inclination is $65.15^{\circ}$, and the $O-C$ residuals
for the $J$ band are reasonably small ($\la 0.01$ mag).  However, if
we take the predicted $V$ light curve for this geometry and compare it
to the simulated $V$ light curve, we find that the amplitude of the
black body $V$ light curve is much {\em smaller} than the amplitude
of the simulated $V$ ({\sc NextGen}) light curve.  Thus we have 
reproduced the same
problem that Shahbaz et al.\ (\cite{shahbaz97}) and 
Belczy\'nski \& Mikolajewska (\cite{belczynski98})
had when they attempted simultaneous optical and infrared fits.
Not surprisingly, 
if we attempt to fit both our simulated $V$ and $J$ light curves 
simultaneously using the inclination as the only free parameter,  
the fits to the $J$ light curve get worse (middle panel).
Finally, we fit both light curves simultaneously using both the
inclination and mass ratio as free parameters.  The $O-C$ residuals
for both filters are not excessively large and might be comparable to
the observational errors in the real binary.  
However, the fitted parameters
are quite different from the input ones:  $Q_{\rm fit}=2.41$ and
$i_{\rm fit}=67.2^{\circ}$, compared to input values of
$Q_{\rm inp}=1.67$ and
$i_{\rm inp}=60.0^{\circ}$.  Naturally, the derived component masses
from the fit are significantly different from the input masses:
$M_{\rm giant}({\rm fit})=0.32\,M_{\odot}$ and 
$M_{\rm comp}({\rm fit})=0.78\,M_{\odot}$, compared to input values of
$M_{\rm giant}({\rm inp})=0.71\,M_{\odot}$ and 
$M_{\rm comp}({\rm inp})=1.18\,M_{\odot}$.

\subsection{Rotational broadening kernels}

Measurements of the rotational velocities of the stars in close binaries
provide powerful constraints on the light curve solution.  Indeed,
any light curve solution that specifies the mass ratio, inclination, and
the angular velocity ratios predicts specific values for the
observed values of $V_{\rm rot}\sin i$, provided of course that
at least one radial velocity curve is available.  In cases where a star
fills its critical lobe exactly and is in synchronous rotation
(generally safe assumptions in cataclysmic variables and low mass
X-ray binaries), a measurement of its rotational velocity constrains
the mass ratio of the binary (e.g.\ Wade \& Horne \cite{wade88}).

If spectra with high resolution and high signal-to-noise are available,
then one can use Fourier techniques  to measure the rotational
velocity (e.g.\ Gray \cite{gray92} and cited references).
When only spectra of lower quality are available, then  
often one measures $V_{\rm rot}\sin i$  by comparing the spectrum
of interest with a
spectrum 
of a slowly rotating template star
(observed with similar instrumentation) that has been convolved with 
a broadening kernel $G(\lambda$).  Various trial values of the
width of the broadening kernel are tried until the optimal match is found.
Gray (\cite{gray92}, pp.\ 370-374) 
gives a clear description of how
to compute the broadening kernel $G(\lambda)$ analytically for
the case where the intrinsic line profile $H(\lambda)$
has the same shape over the entire disk.
Essentially,
one can place an $x$,$y$ coordinate system on 
the apparent disk of the star on the sky where the $y$-axis is parallel
to the axis of rotation.  The disk of the star 
can then be divided up into
a number of strips  parallel to the $y$-axis, 
each having its own Doppler shift according to its $x$-coordinate.
The broadening kernel is evaluated at a particular Doppler shift by
integrating the intensity over the appropriate strip.
Normally one assumes a linear limb darkening law with a coefficient
of $x=0.6$ when computing $G(\lambda)$.
Finally, the broadened line profile is the convolution of the intrinsic
line profile with the kernel: $H_{\rm rot}(\lambda)=G(\lambda)*H(\lambda)$.

\begin{figure}
\centering
\centerline{\epsfxsize=8.2cm
\epsfbox{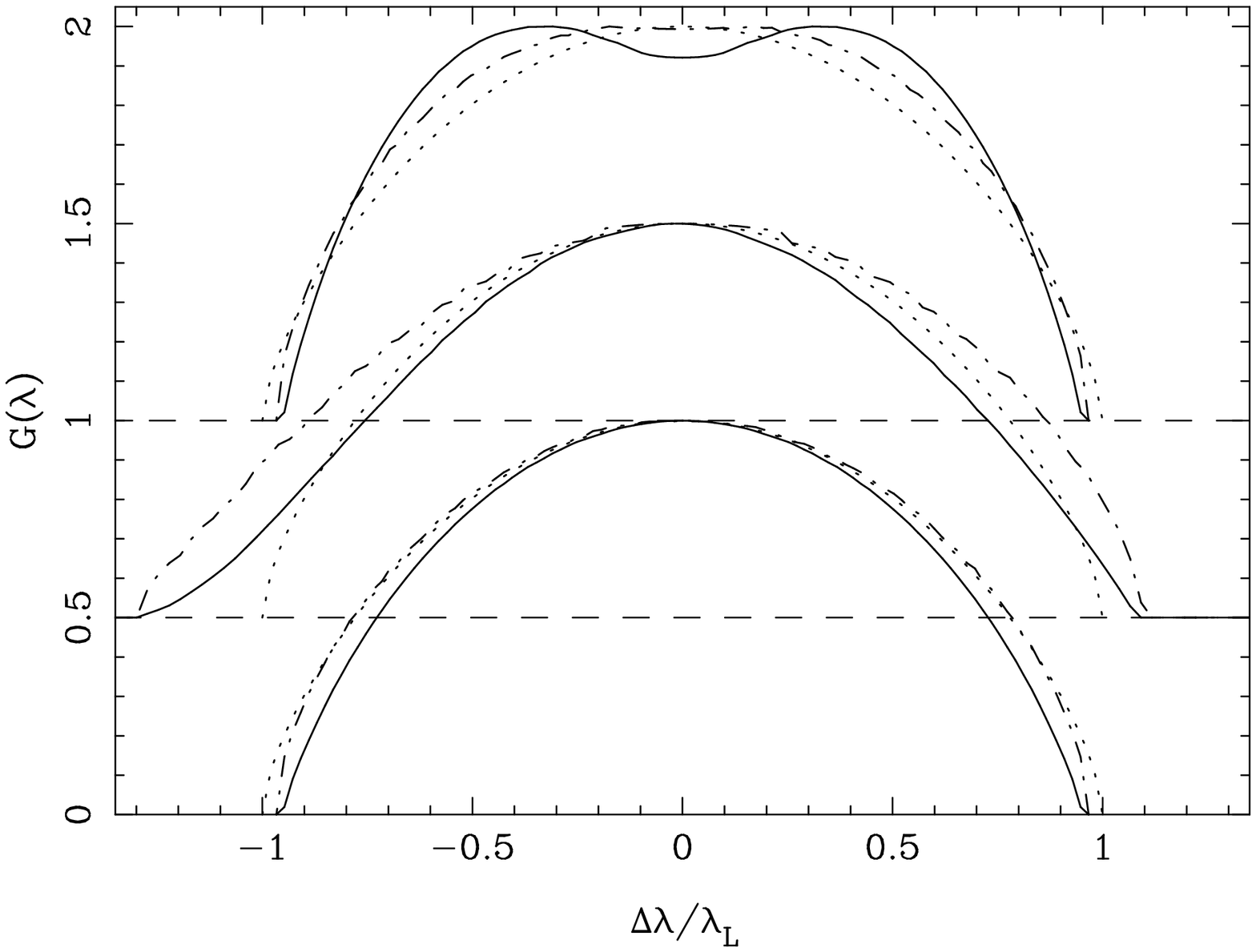}}
\caption{Rotational broadening kernels $G(\lambda)$
for the star described in the caption to Fig.\ \protect\ref{fig4}
for three different phases: 0.0 (lower curves), $90^{\circ}$ (middle
curves, offset by 0.5), and $180^{\circ}$ (upper curves,
offset by 1.0) (the giant is behind its invisible companion at
phase $180^{\circ}$).  The solid lines are the kernels for
Roche geometry and {\sc NextGen} intensities.  The dash-dotted
curves are for Roche geometry and black body intensities plus a linear
limb darkening law with a coefficient of $x=0.6$.  The dotted
curves are analytic kernels with a limb darkening coefficient of 0.6.
}
\label{fig6}
\end{figure}

\begin{figure}
\centering
\centerline{\epsfxsize=8.2cm
\epsfbox{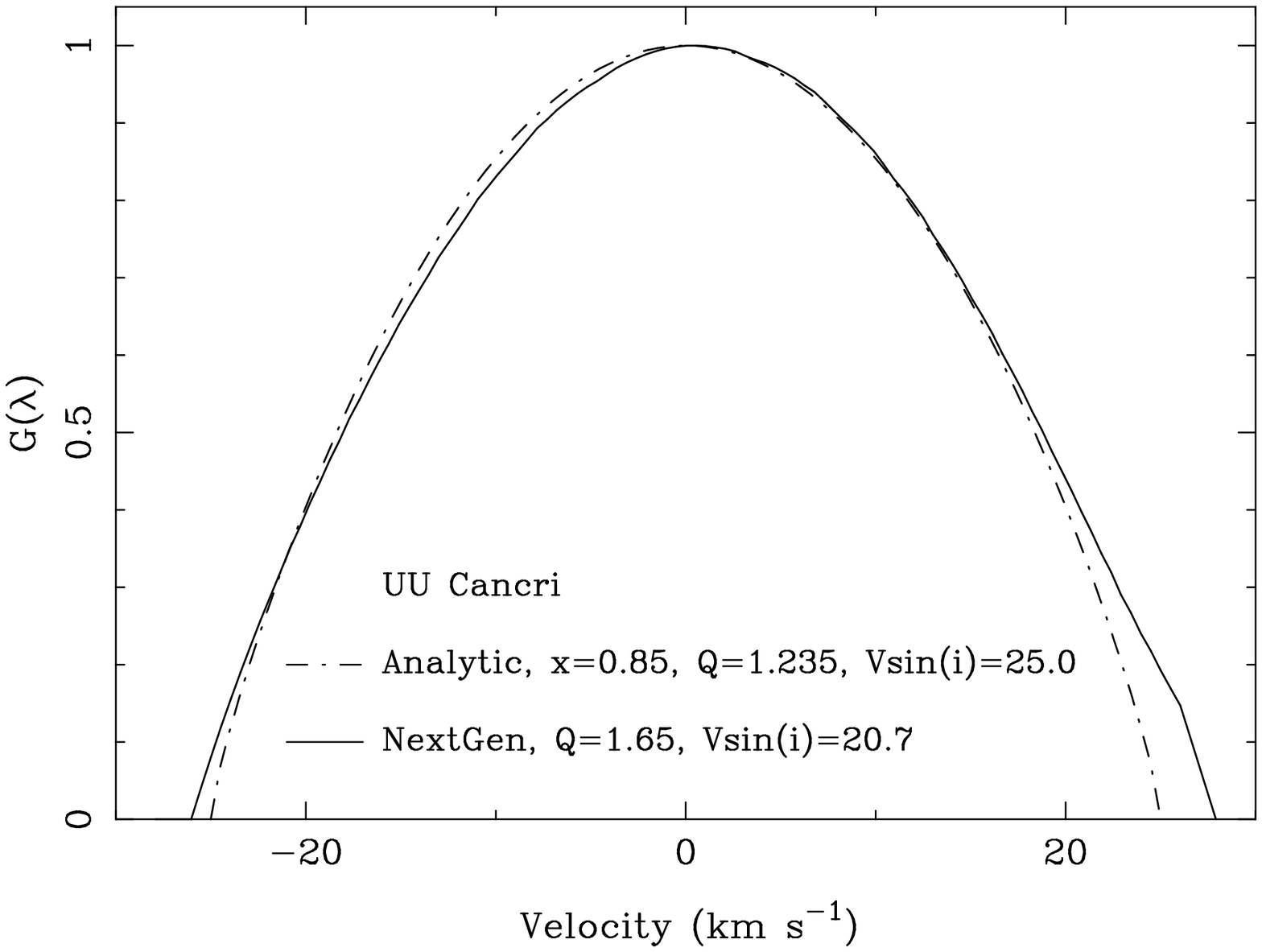}}
\caption{Rotational broadening kernels $G(\lambda)$
for 
\protect\object{UU Cancri}.  The solid line is the kernel computed for
a mass ratio of $Q=1.65$, which corresponds to a ``mean'' rotational
velocity of $2\pi R_{\rm eff}/P=20.7$ km s$^{-1}$.
The dash-dotted line is the analytic kernel computed by
Eaton et al.\ (\protect\cite{eaton91}), corresponding to 
$Q=1.235$ and
$2\pi R_{\rm eff}/P=25.0$ km s$^{-1}$.
}
\label{fig7}
\end{figure}

The above discussion applies to spherical stars.
A star that fills a substantial fraction of its Roche lobe departs markedly
from spherical symmetry, and as such will have distorted line
profiles (e.g.\ Kopal \cite{kopal59}; 
Marsh et al.\ \cite{marsh94}; Shahbaz \cite{shahbaz98}).
Furthermore, the degree of the line profile distortion depends on the
Roche geometry (i.e.\ mass ratio and inclination) and the orbital phase
(Shahbaz \cite{shahbaz98}).
The intensity maps that the ELC program can write (e.g.\ Fig.\ \ref{fig2})
can be used to numerically evaluate $G(\lambda$), thereby accounting
for both the nonspherical shape of the star and the nonlinear limb darkening.
Fig.\ \ref{fig6} shows 
rotational broadening kernels $G(\lambda)$
for the star described in the caption to Fig.\ \ref{fig4}
for phases 0.0, $90^{\circ}$, and $180^{\circ}$,
when the giant is behind its invisible companion.
The situation is complicated.  Near phase 0.0, the projected surface of
the giant in the plane of the sky is not too different from a circle (it
is flattened slightly in the $y$-direction).  Thus the observed profile
will be nearly symmetric.  However, the kernel, computed using the
{\sc NextGen} intensities, is {\em narrower} than the
analytic kernel and the kernel computed using black body intensities
and linear limb darkening since the 
star with the {\sc NextGen} intensities appears smaller on the sky
(Fig.\ \ref{fig2}).
At phase $90^{\circ}$ (quadrature), the star is elongated in the
$x$-direction.  As a result, the broadening kernel computed
from the black body intensities is {\em wider} than the analytic
kernel, and it is asymmetric.  As a result of the sharp cut-off in
the limb darkening,  
the kernel computed using the {\sc NextGen} intensities
in narrower than the black body kernel.  
However, for this
particular situation, the {\sc NextGen} kernel has roughly
the same width as the analytic kernel.
Finally, at phase $180^{\circ}$, when the giant is behind its
unseen companion, all three kernels have more or less the same full
width at half maximum.  However, 
the kernel computed using the black body intensities
is flat-topped
(Shahbaz (\cite{shahbaz98}) also noted absorption line profiles
with flat bottoms near this phase (his Fig.\ 1)), and the kernel
computed with the {\sc NextGen} intensities has a local minimum
at zero velocity.  The reason the kernels have flat tops or even  central
depressions near
phase $180^{\circ}$ is quite straightforward.  The $L_1$ point is in full
view near this phase, and for high inclinations there are no points on
the apparent stellar disk
with $\mu=1$.  For the example shown in Fig.\ \ref{fig6}, $\mu<0.924$ 
everywhere on the star at
phase $180^{\circ}$.  Hence the central parts of the apparent stellar
disk are fainter  (owing to limb darkening)
than they would have been 
in the spherical case and there is less 
contribution to the kernel near zero velocity.  
If the intensities are computed using
the {\sc NextGen} table and the gravity is sufficiently low, then the
central part of the star can be even darker.  In some cases such as the
one shown in Fig.\ \ref{fig6},
the $L_1$ point
will be so dark that the broadening kernel will actually have
a central depression near zero velocity.

Given the variety of distortions in the broadening kernel 
as shown in Fig.\ \ref{fig6}, it would be prudent to compute
phase-specific broadening kernels when extracting $V_{\rm rot}\sin
i$  measurements from spectra (see also Shahbaz \cite{shahbaz98}).
As a specific example, we consider the eclipsing
binary \object{UU Cancri}, which  consists
of a K giant in a 96.7 day orbit about an essentially invisible 
(in the optical) companion
(e.g.\ Eaton et al.\ \cite{eaton91}).  
There is good evidence
for an accretion disk around the unseen companion (Zola et al.\
\cite{zola94}), so we will assume that the K giant fills its
Roche lobe exactly and that it is in synchronous rotation.
Eaton et al.\ (\cite{eaton91}) 
obtained a series of high resolution spectra
and noted a change in the Doppler broadening of certain metallic 
absorption lines as a function of phase. This behaviour is consistent
with expectations (Shahbaz \cite{shahbaz98}).  
Using a linear limb darkening law with
$x=0.85$, Eaton et al.\ (\cite{eaton91}) determined
a rotational velocity of 	$V_{\rm rot}\sin i=25\pm 1$
km s$^{-1}$ for a spectrum near a quadrature phase
and derived a mass ratio of $Q=M_{\rm comp}/M_{\rm giant}\approx
1.2$ (using ELC, we derive a mass ratio $Q=1.235$ from 
$V_{\rm rot}\sin i=25\pm 1$, 
where we have adopted a mass
function for the unseen companion of $f(M)=0.56\,M_{\odot}$
(Popper \cite{popper77})).  
We plot in Fig.\ \ref{fig7}
the analytic broadening kernel for $x=0.85$
and a mean rotational velocity of
$2\pi R_{\rm eff}/P=25$ km s$^{-1}$, where $R_{\rm eff}$ is the 
sphere-equivalent Roche lobe radius.  We also show a broadening kernel
computed using {\sc NextGen} intensities, assuming $Q=1.65$ and
$2\pi R_{\rm eff}/P=20.7$ km s$^{-1}$.  We have adopted an inclination 
of $i=89.6^{\circ}$ and an effective temperature of 3900~K
for the K giant
(Zola et al.\ \cite{zola94}).  
The two broadening kernels have nearly the
same full width at half maximum.  However, the implied component
masses for the two cases are quite different.  Assuming $i\approx 90^{\circ}$,
$Q=1.235$ (the value from Eaton et al.\ \cite{eaton91}) implies
$M_{\rm giant}=1.49\,M_{\odot}$,
$M_{\rm comp}=1.83\,M_{\odot}$,
and
$\log g_{\rm giant}=1.25$, whereas our value
of $Q=1.65$ gives
$M_{\rm giant}=0.88\,M_{\odot}$,
$M_{\rm comp}=1.44\,M_{\odot}$,
and
$\log g_{\rm giant}=1.19$.  Zola et al.\ (\cite{zola94}) 
find a {photometric}
mass ratio from their light curve solutions of
$q=M_{\rm giant}/M_{\rm comp}=0.564\pm 0.006$, which in our
notation is $Q=1.77\pm 0.01$.  
The case of 
\object{UU Cnc} is perhaps an extreme example, but it serves to illustrate
the importance of using broadening kernels which account for
deviations from linear limb darkening and spherical geometry.

\subsection{Accuracy of the integration and interpolation}

\begin{figure}
\centering
\centerline{\epsfxsize=8.2cm
\epsfbox{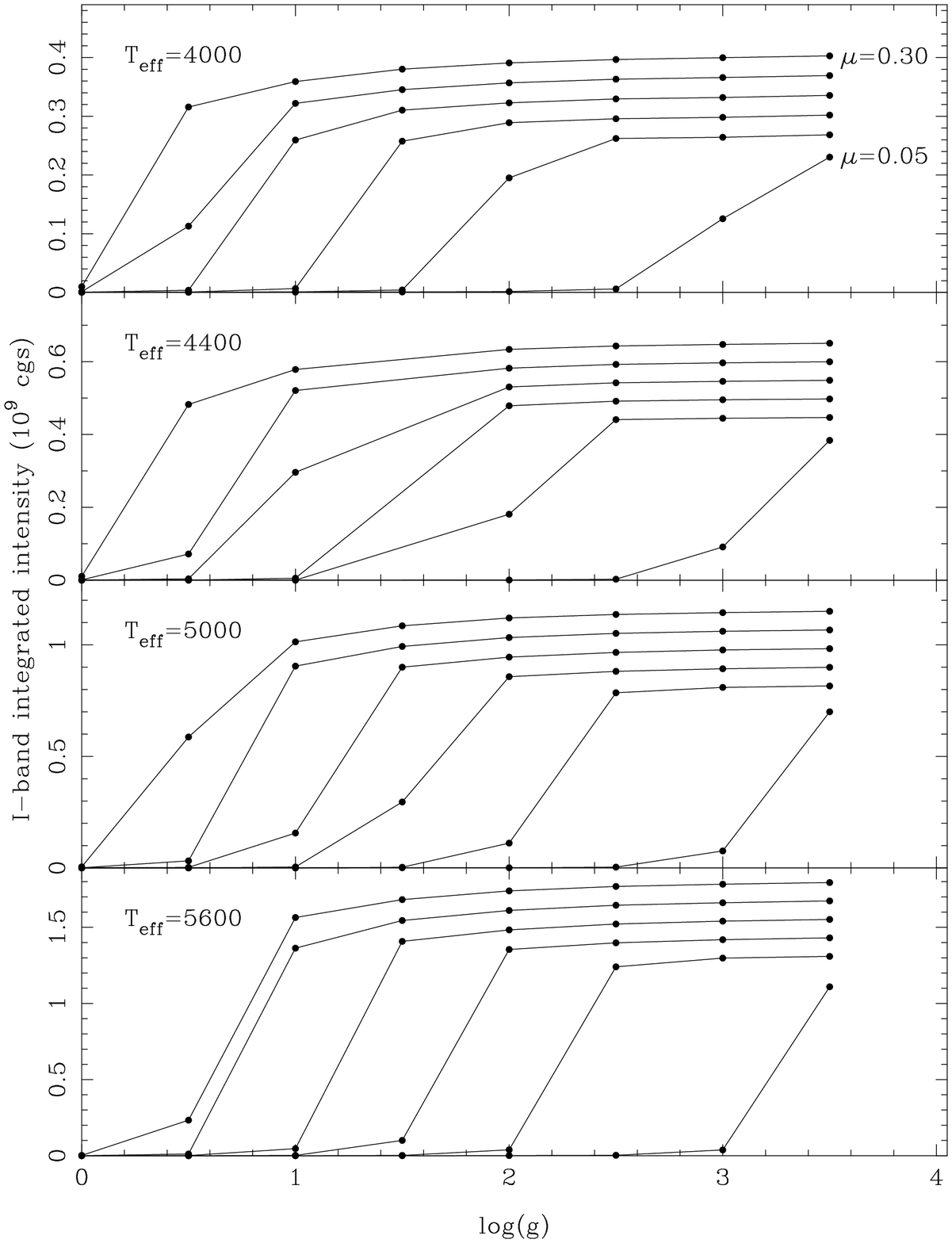}}
\caption{The curves show the
integrated $I$-band specific intensity for various
specific angles and effective temperatures as a function of
the gravity.  The intensity is generally a smooth function
for a fixed temperature and emergent angle.  Thus a linear interpolation
in $\log g$ is reasonably accurate.
}
\label{fig8}
\end{figure}

\begin{figure}
\centering
\centerline{\epsfxsize=8.2cm
\epsfbox{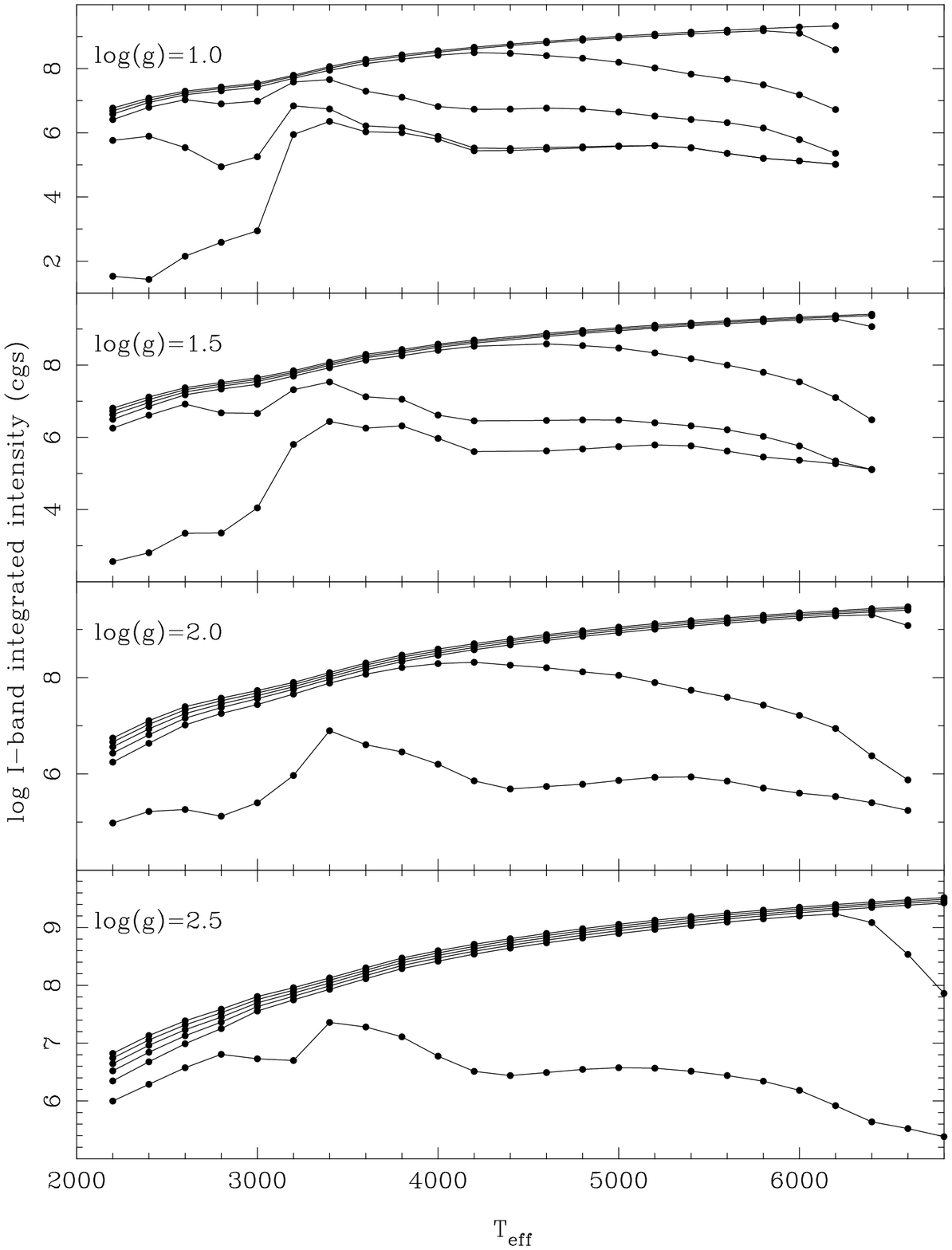}}
\caption{Similar to Fig.\
\protect\ref{fig8}, except the intensity is
now displayed as a function of the effective temperature.  Note
the $y$-axis scale here is logarithmic.  Within each panel, the
specific angles are from bottom to top $\mu=0.05$,
$0.10$,
$0.15$,
$0.20$,
$0.25$, and
$0.30$.  In most cases the $\mu=0.25$ and the $\mu=0.30$ curves are
nearly the same.  As in Fig.\
\protect\ref{fig8}, the  curves are quite
smooth, and a linear interpolation of the intensity in
$T_{\rm eff}$ is reasonably accurate.
}
\label{fig9}
\end{figure}

\begin{figure}
\centering
\centerline{\epsfxsize=8.2cm
\epsfbox{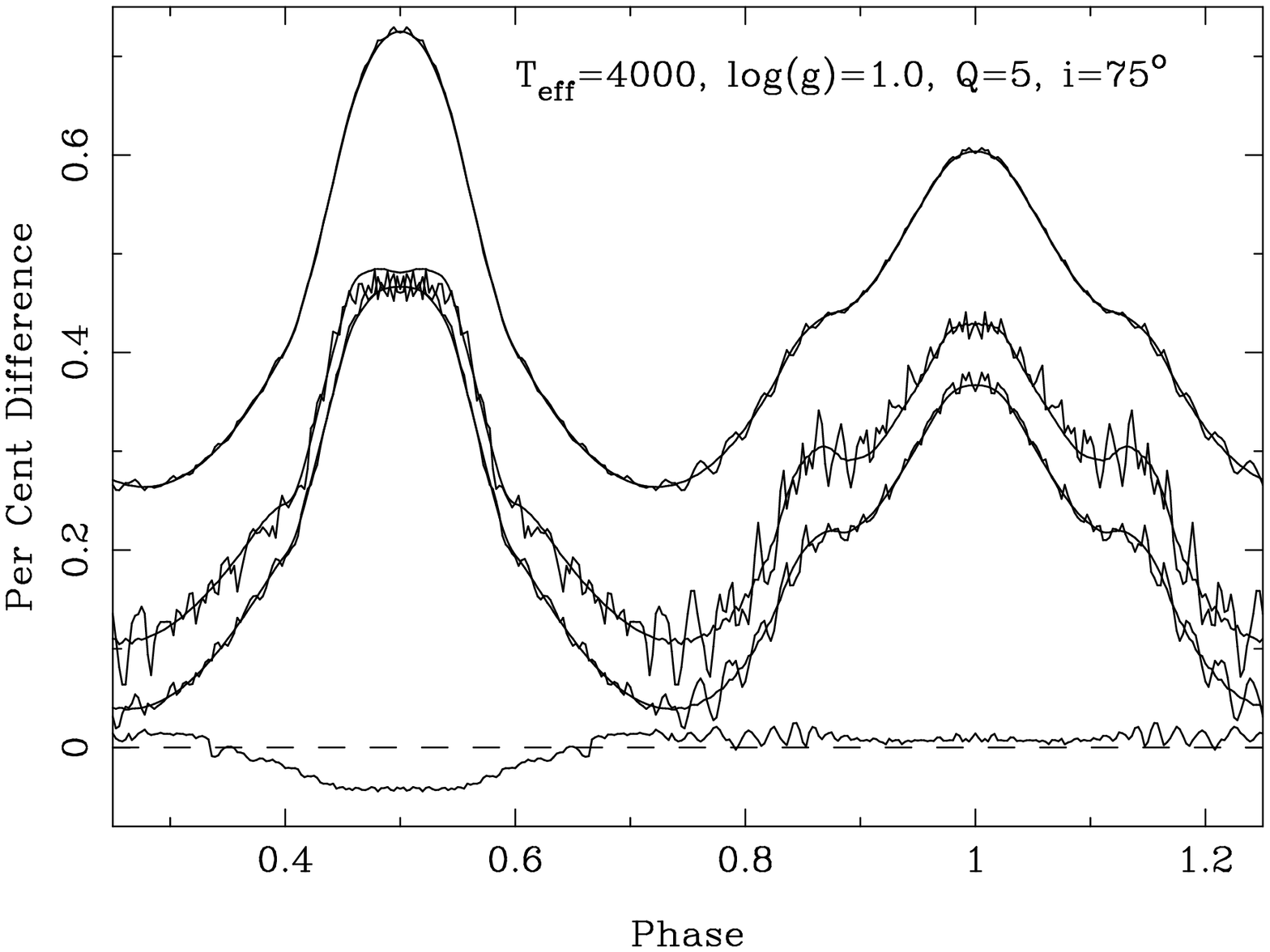}}
\caption{``Difference'' plots in the light curves for the star
described in the caption to Fig.\ \protect\ref{fig4}.  
The top three sets of curves are difference curves for
$V$, $K$, and $I$.  For each filter, there are two curves corresponding
to a ``standard'' grid of surface elements ($N_{\alpha}=40$, $N_{\beta}
=14$) and a ``fine'' grid  
($N_{\alpha}=120$, $N_{\beta}
=40$).  The curves for the fine grids are smoother.  The curve at the
bottom compares the regular $V$ models computed using the two different
grids.
}
\label{fig10}
\end{figure}

Given the relatively large differences in the predicted light curves
and broadening kernels that we predict, 
it is worthwhile to discuss the numerical accuracy of the output
light curves.  
Since the computation of the {\sc NextGen} models is somewhat
time consuming, we cannot tabulate models for all effective
temperatures and gravities.  We have done extensive testing
on the integration and interpolation procedures, and we
believe  the present table
(steps of 0.5 dex in $\log g$ and 200~K in
$T_{\rm eff}$) is a reasonable compromise between
CPU time and sampling accuracy.  There are various ways to see
how ``smooth'' the filter-integrated intensities are as a function of
the effective temperature and gravity.  Figs.\ \ref{fig8} and
\ref{fig9} show two such representations.  In Fig.\
\ref{fig8} we show as a function of
the gravity the integrated $I$-band intensity for four
effective temperatures (4000, 4400, 5000, and 5600~K)
at six different emergent angles ($\mu=0.05$, 0.10, 0.15, 0.20, 0.25,
and 0.30). In general, these curves are smooth, and a linear interpolation
of the intensity in the $\log g$ direction gives reasonable
accuracy.  Fig.\ \ref{fig9}
shows a similar plot, but with the intensities as a function
of the effective temperature.  As before, the curves are quite smooth
and a linear interpolation in $T_{\rm eff}$ gives reasonable
results.  There is some irregular behaviour in the curves for
$\mu=0.05$ and $\mu=0.10$.  In practice, however, the contribution to
the overall integrated light curve from angles less than 0.1 is small
since the intensities are weighted by the value of $\mu$.

Another way to test the accuracy of the interpolation is to leave 
specific models out of the table, 
compute a light curve, and compare it to the light curve computed using
the full intensity table.
Fig.\ 
\ref{fig10} shows results of this exercise for the star described in
the caption of Fig.\ \ref{fig4}.  We computed ``regular'' light curves
where we used all models in the intensity table and ``cut'' light
curves where the $T_{\rm eff}=4000$, $\log g=1.0$ model was excluded
from the table.  Thus, at $T_{\rm eff}=4000$, the local step size in
$\log g$ was 1 dex, rather than 0.5 dex.
To compare the light curves, we define the
``difference curve''
as $D=100(I_{\rm reg}-I_{\rm cut})/I_{\rm reg}$,
where $I_{\rm reg}$ is the integrated light computed using the full
intensity table and $I_{\rm cut}$ is the integrated light computed using
the intensity table with the $T_{\rm eff}=4000$, $\log g=1.0$
model removed.  We see that the systematic
difference between the two curves is usually about 0.3
per cent
and is at most
0.75
per  cent in $V$.  The corresponding values for $I$ and $K$ are even smaller.
We also computed light curves using two different grids of surface
elements:  the ``standard'' grid with $N_{\alpha}=40$ and $N_{\beta}=14$
and the ``fine'' grid with $N_{\alpha}=120$ and $N_{\beta}=40$.  
Using the fine grid has little effect on the difference curves:
the curves are smoother but the overall shapes are the same.  The lowermost
curve in Fig.\ \ref{fig10} compares the regular $V$ light curve
computed using the standard and fine grids.  The two curves are the same
to within 0.05 per cent
(i.e.\ less than about 0.5 millimag).  We conclude that for most
situations our regular intensity table and the standard grid are
adequate.

There is nothing in our integration
and interpolation techniques that limits us to $\approx 1$ per cent
accuracy.  Since our interpolation scheme does not have a fixed
step size in the temperature or gravity we
can easily add a few models with the appropriate temperatures
and gravities for special cases where the data demand the highest accuracy.
Other light curve codes also suffer systematic errors on the few
per cent level.  For example, most versions of the W-D code
use a single limb darkening law for the entire star.  Since the temperature
and gravity are not constant over the surface of a tidally distorted
star,
small systematic errors can be present when a limb darkening law for
a single temperature and gravity are used.  Van Hamme \& Wilson
(\cite{vanhamme94}) have discussed this point in more detail.

\section{Discussion and summary}

In this paper we have presented a way to include specific intensities
from detailed model atmosphere computations into a light curve synthesis
code.  We have shown that using the model atmosphere intensities directly
is almost required for cool giants since the limb darkening behaviour
for a cool low-gravity atmosphere is radically different than simple
one- or two-parameter limb darkening laws (Fig.\ \ref{fig1}).
This departure from the near-linear limb darkening law is a consequence
of the spherical geometry used in the computation of the {\sc NextGen}
models.
Other workers have computed spherical model atmospheres for
cool giants before (e.g.\
Scholz \& Tsuji \cite{scholz84}; 
Scholz \& Takeda \cite{scholz87};
Plez et al.\ \cite{plez92};
J{\o}rgensen et al.\ \cite{jorgensen92}), 
and the strongly nonlinear
limb darkening behaviour has previously been noted by
Scholz \& Takeda (\cite{scholz87}), 
and more recently by Hofmann \& Scholz (\cite{hofmann98}).  
We have shown that
this strongly nonlinear limb darkening has a large effect on the light
curve amplitude.  In general, the light curves computed using the
{\sc NextGen} intensities have larger amplitudes than the light curves
computed using the same geometry but with  black body intensities and
a one- or two-parameter limb darkening law.  Also, 
the strongly nonlinear
limb darkening has an effect on rotational broadening kernels
in that for spherical stars the kernel computed using the {\sc
NextGen} intensities will be narrower than a numerical or an analytic
kernel that has a linear limb darkening law.  
If the star fills a substantial fraction of its Roche lobe, then the
broadening kernel will also be different than the analytic one
owing to the nonspherical shape of the star.

Thus we basically have two different types of models with which to fit 
close binary light curves:  our ELC code with the {\sc NextGen}
intensities, and codes that use either black body intensities or
plane-parallel model atmosphere intensities and near-linear limb darkening
(e.g.\ W-D, ELC in black body mode, ELC with the Kurucz table, etc.).
It should be possible to test which class of models provides the better
description of the observational data available for binaries with cool
giants.  In this regard, we have shown that observations in several bandpasses
are useful in discriminating between the models. For example,
Figs.\ \ref{fig4} and \ref{rzscuti}
show how the difference in the depths of the minima between the
{\sc NextGen} and Kurucz light curves depends on the bandpass, while
Fig.\ \ref{fitfig} shows that black body light curves provide a poor
simultaneous fit to {\sc NextGen} $V$ and $J$ light curves of
a binary like 
\object{T CrB}.  There is no shortage of potentially good
binary stars to study.  For example,
\object{T CrB}, as we have already noted, has $V$, $I$,
$J$, and $K$ light curves available (although the light curves are perhaps
somewhat noisy).  
\object{RZ Ophiuchi} is an eclipsing binary consisting of
a hot $\approx$B star and a K supergiant.  The orbital period
is 262 days.  Reasonably good light curves in the bluer filters exist
(e.g.\ Knee et al.\ \cite{knee86}; Olson \cite{olson93}), 
and additional observations
in $J$, $H$, and $K$ would be extremely valuable.
Kruszewski \& Semeniuk (\cite{kruszewski99}) 
present a catalog of poorly studied eclipsing
binaries with good parallaxes measured by the {\em Hipparcos} mission.
Many of these binaries have long periods (more than 10 days) and contain
evolved cool components.   Needless to say, we encourage observers to 
systematically study these binaries
in multiple colours.
With a binary of known distance, one has the
added challenge of obtaining the correct integrated flux.  

Our interest should not be confined to binaries with a supergiant
component.  For example, there appears to be a small class of 
``long period'' Algol type binaries (periods between about 10 and 20
days) where the mass losing cool star appears to be slightly underfilling
its Roche lobe.  Some examples are \object{WW Andromedae}
(Olson \& Etzel \cite{olson93a}), 
\object{S Cancri} (Olson \& Etzel \cite{olson93b}), and 
\object{DN Orionis}
(Etzel \& Olson \cite{etzel95}).  The cool stars in these three systems 
have surface gravities near $\log g\approx 2.5$, so the sphericity effects
are not as pronounced as they are in systems like 
\object{T CrB}.  On the other hand,
all three of these binaries have excellent five-colour ($I$(Kron)$ybvu$)
photometry, and 
\object{S Cnc} and 
\object{DN Ori} are totally eclipsing and are double-lined.
Olson and Etzel report that in each case, the cool star seems to underfill
its Roche lobe by about 10 per cent.  It would be worthwhile to 
re-evaluate the light curve solutions for these three stars using ELC and
the {\sc NextGen} intensities to see if the sphericity effects can account
for the apparent slight underfilling of the Roche lobes by the cool stars.

\begin{acknowledgements}
We thank Marten van Kerkwijk, Frank Verbunt, and Robert E. Wilson
for helpful discussions.  JAO gratefully acknowledges the hospitality
of the Institute of Astronomy in Cambridge.       
PHH's work was supported in part by the 
CNRS, INSU and by NSF grant
AST-9720704, NASA ATP grant NAG 5-8425 and LTSA grant NAG 5-3619, as
well as NASA/JPL grant 961582 to the University of Georgia.
This work was supported in part by the P\^ole
Scientifique de Mod\'elisation Num\'erique at ENS-Lyon.  Some of the
calculations presented in this paper were performed on 
the IBM SP and SGI Origin 2000 of the UGA UCNS, on the IBM ``Blue Horizon'' of
the San Diego Supercomputer Center (SDSC), with support from the
National Science Foundation, and on the Cray T3E and IBM SP of the NERSC with
support from the DoE.  We thank all these institutions for a generous
allocation of computer time.
\end{acknowledgements}

\appendix

\section{Brief outline of the ELC code}

\subsection{Introduction and history}

Yorham Avni was interested in, among other things, mass determinations of the 
compact objects in high mass X-ray binaries (e.g.\ 
\object{Cyg X-1}, \object{Cen X-3}, etc.).
During the course of his research 
he wrote a FORTRAN code to compute the ellipsoidal light curve of a single
(usually) Roche-lobe filling star 
(Avni \& Bahcall \cite{avni75}; Avni \cite{avni78}).
Avni passed on this code to Jeff McClintock and Ron Remillard shortly before
his death in March of 1988 
(see McClintock \& Remillard \cite{mcclintock90}).  The code
was passed on to JAO from
Ron Remillard sometime in 1994 when JAO was a graduate
student at Yale.  During the spring of 1995 it became clear that 
the black hole binary 
\object{GRO J1655-40} was an eclipsing system and that the
Avni code in its original form was not adequate.  The code was
substantially
modified by JAO in the summer of 1995 to include light from an accretion
disk and to account for eclipses (Orosz \& Bailyn \cite{orosz97}).

The code described in 
Orosz \& Bailyn (\cite{orosz97}) 
is somewhat unwieldy and it was becoming more
and more
difficult to read and modify.  Therefore a more
general and  more
modular code was developed by JAO.  
Although we have used much of Avni's
notation, and we have followed his basic method of setting up the Roche
geometry and integrating the observable flux, most of the code 
is new.  Owing to space limitations, we cannot give below each equation
used.  Rather, we give below some details of the parts of the
code which have been substantially revised with respect to the earlier
versions (there are numerous other papers and
texts which go into varying amounts
of detail,  
Wilson \& Devinney \cite{wilson71}; 
Wilson \cite{wilson79}; 
Avni \& Bahcall \cite{avni75}; 
Avni \cite{avni78}; 
Linnell \cite{linnell84};
Wilson \cite{wilson90};
Orosz \& Bailyn \cite{orosz97}).

\subsection{The Potential}

We assume surface of the star is an equipotential surface
of the following potential, which includes the gravitational, centrifugal,
and Coriolis forces (Avni \& Bahcall \cite{avni75}):
\begin{equation}
\Psi={GM_1\over D}\left[{1\over r_1}+{Q\over r_2}-Qx
+{1+Q\over 2}\left({\omega_1\over \omega_{\rm orb}}\right)^2
(x^2+y^2)\right]
\label{pot}
\end{equation}
where $M_1$ is the mass of the star under consideration, $D$ is separation
between two stars, $Q=M_2/M_1$ is the mass ratio, 
$\omega_1$ is the star's rotational angular velocity,
$\omega_{\rm orb}$ is the orbital angular velocity, $r_1$ and
$r_2$ are the distances to the stellar centres in units of $D$,
and $x$ and $y$ are normalised coordinates centred at star 1.  
For a binary with a given mass ratio $Q$, rotational angular
velocity $\omega_1$ and
orbital angular velocity $\omega_{\rm orb}$, there is a critical value of
the potential $\Psi_{\rm crit}$ where the star exactly fills its limiting
lobe.  Stars that are smaller then their limiting lobes will have
$\Psi > \Psi_{\rm crit}$.  

To fully define the surface of the star, the user specifies
$Q$, $\Omega=\omega_1/\omega_{\rm orb}$, and the ``filling factor''
$f\equiv x_{\rm point}/x_{L1}$, where
$x_{\rm point}$ is the $x$-coordinate of the ``nose'' of the star
and $x_{L1}$ is the $x$-coordinate of the $L_1$ point.
In our case, the filling factor $f$ 
is exactly 1 for Roche lobe-filling stars and less than 1
for detached stars. Situations with $f>1$ (the contact binaries) are
currently not allowed.  
When $f<1$, the program computes 
$x_{\rm point}$ from $x_{\rm point}=fx_{L1}$ and then computes 
$\Psi(x_{\rm point},0,0)$, which is the adopted potential for the star.
Once the surface of the star is defined, a grid
of surface elements is made using a 
polar coordinate system with
$N_{\alpha}$ latitude rows equally spaced in the angle $\theta$ and
$4*N_{\beta}$ longitude points per latitude row equally spaced in
the angle $\phi$.   (Earlier versions of the code used a cylindrical
coordinate system with 
$N_{\alpha}$ rings equally spaced
along the line of centres, running from the $L_1$ point to the back
of the star and $4*N_{\beta}$ surface points per ring equally spaced
in angle.)  
It is convenient to use an internal rectangular coordinate system
centred on star 1
where the $x$ axis points to the centre of the other object and
the $z$ axis is parallel to the direction of $\omega_1$. 
The value of $\Psi$ and its derivatives
are computed for each element, and from these quantities the local
gravities $g(x,y,z)$
and the surface normal vectors follow.

\subsection{Mean temperature vs.\ polar temperature}

The temperature of the secondary star was defined in the 
Avni code by its polar temperature $T_{\rm pole}$.  Given
$T_{\rm pole}$, the temperatures of the other surface elements followed
from  the well-known von Zeipel relation:
\begin{equation}
{T(x,y,z)\over T_{\rm pole}}=\left[{g(x,y,z)\over g_{\rm pole}}
\right]^{\beta}.
\end{equation}
The exponent $\beta$ is 0.25 for stars with a radiative atmosphere
(von Zeipel \cite{vonzeipel24}) 
and 0.08 for stars with convective envelopes
(Lucy \cite{lucy67}).  
Wilson (\cite{wilson79}) has pointed out that the polar temperature
and the {\em mean} temperature of a distorted star
will be different.  In most cases, of course, one measures the
mean temperature via the spectral type or colour index.
Therefore, following Wilson (\cite{wilson79}), we now
have as input the mean (or equivalently effective)
temperature  of the star,
denoted by $T_{\rm eff}$.  The effective temperature is computed
from the bolometric luminosity
\begin{equation}
L=\sigma S T_{\rm eff}^4
\end{equation}
where $S$ is the surface area.  $T_{\rm pole}$ is then given by
(Wilson 1979)
\begin{equation}
T_{\rm pole}=T_{\rm eff}\left[S\over \int_{\rm surface}
g_{\rm norm}(x,y,z)^{4\beta}dS(x,y,z)\right]^{1/4}
\end{equation}
where $g_{\rm norm}(x,y,z)=g(x,y,z)/g_{\rm pole}$ and 
$dS(x,y,z)$ is an element of surface area (Eqs.\ (A8), (A9), and
(A10) of Orosz \& Bailyn \cite{orosz97}).

\subsection{Addition of a second star}

Adding a second star to the code is relatively simple.  We ``flip'' the
mass ratio (define $Q^{\prime}=1/Q$), and solve for the potential and
its gradients using the same subroutines as for the first star.  When
integrating the observed flux, we add $180^{\circ}$ to the phase
and use the same subroutines as for the first star.  One complication
occurs when the second star is not in synchronous rotation (Wilson
\cite{wilson79}).  
In this case, the $x$ derivative of the potential that is used
in the detailed reflection routine (see below)
has
a different form.  Thus the subroutine that returns the potential
gradients returns two sets of $x$ derivatives and the appropriate
one is used for the detailed reflection.

\subsection{Detailed reflection scheme}

Stars in a close binary can heat each other, and this mutual heating
leads to easily observed consequences.
Wilson (\cite{wilson90}) 
divides the ``reflection'' theory into four main parts:
the geometric aspect, the bolometric energy exchange,
the intensity from an irradiated stellar atmosphere, and 
the effect on the envelope structure.  In this paper Wilson
gives a complete description of his ``detailed reflection'' scheme
which treats the first two parts of the theory essentially exactly.
We have fully implemented Wilson's treatment of the reflection effect,
which is a big improvement over the scheme used by Orosz \& Bailyn
(\cite{orosz97}).  There are two points about this scheme that are noteworthy:

(i)~~Wilson's scheme makes use of {\em bolometric} limb darkening
approximations.  We have seen that the limb darkening in cool giants
is not well parameterised by the common limb darkening laws.  
Fortunately, in most practical situations, a relatively hot star with a 
high surface gravity
($\log g > 4$) 
irradiates
a much cooler star.  The irradiation of the hot star by the cool star
can be neglected (in which case we don't care about the details of
the cool star limb darkening), and
the well-known limb darkening laws apply nicely to
the hot, high-gravity star.  

(ii)~Wilson introduced the use of the $R$-function in the reflection scheme
to allow for multiple reflection.  For each element, the $R$-function is
defined by
\begin{eqnarray}
R_1(x,y,z) &=& 1+{F^{\prime}_2\over F_1(x,y,z)} \\
R_2(x,y,z) &=& 1+{F^{\prime}_1\over F_2(x,y,z)} 
\end{eqnarray}
where ${F^{\prime}_2/F_1(x,y,z)}$ is the ratio of the total
irradiating flux from star 2 seen at the local surface element on
star 1 and ${F^{\prime}_1/F_2(x,y,z)}$ is the reverse.
The new temperatures of the irradiated surface elements are then
\begin{eqnarray}
T_1^{\rm new}(x,y,z) &=& T_1^{\rm old}(x,y,z)R_1^{1/4}(x,y,z) \\
T_2^{\rm new}(x,y,z) &=& T_2^{\rm old}(x,y,z)R_2^{1/4}(x,y,z).
\end{eqnarray}
It is usually assumed that the specific intensity of an irradiated
surface element is the same as the intensity of an unheated surface
element with the same temperature.  At some point the irradiation become
intense enough that this assumption must break down.  Alencar \&
Vaz (\cite{alencar99}) 
have computed some irradiated atmosphere models and presented
limb darkening coefficients for use in light curve synthesis codes.
The use of these coefficients is perhaps somewhat limited as the widely
available versions of the popular W-D code have only a single limb
darkening law for the entire star (which of course includes the unheated
back hemisphere).
One of us (PHH) is in the process
of computing irradiated atmospheres with the {\sc PHOENIX} code.  It should
be relatively simple to include the irradiated atmospheres into the ELC code,
at least for specific binaries.  There exist many eclipsing close
binaries where a hot subdwarf O/B star irradiates its G-M main sequence
companion 
(e.g.\ Hilditch et al.\ \cite{hilditch96}).  Such systems can
provide strong tests of irradiated atmospheres.

\subsection{Flared accretion disk}

The accretion disk used in the Orosz \& Bailyn (\cite{orosz97}) 
code was a flattened
cylinder.  To make the disk perhaps more realistic, we have modified the
disk so that its thickness as a function of the radius is proportional
to the radius.  The disk, if present, is always around star 2.  Star
2 does not necessarily have to be present (as in an X-ray binary).
The disk is described by five basic parameters:
$r_{\rm outer}$, the radius of the outer edge of the disk in
terms of the effective Roche lobe radius of star 2;
$r_{\rm inner}$,  the inner radius of the disk in the same units
as the filling factor of star 2 ($f_2$); $\beta_{\rm rim}$, the opening
angle of the disk rim above the plane; $T_{\rm disk}$, the
temperature of the {\em inner} disk (in the Orosz \& Bailyn 
(\cite{orosz97})
code the temperature of the {\em outer} rim  was specified); and
$\xi$, the power-law exponent on the temperature profile
of the disk:
\begin{equation}
T(r)=T_{\rm disk}(r/r_{\rm inner})^{\xi}.
\end{equation}
For a steady-state accretion disk, $\xi=-3/4$ (Pringle \cite{pringle81}).  
For a disk
heated by a central source, the exponent $\xi$  can take on a range
of values
($-3/4\la \xi \la -1/2$ (e.g.\
Friedjung \cite{fried85};
Vrtilek et al.\ \cite{vrtilek90};
Bell \cite{bell99})).

Since the surface of the disk is flared, each element on the face of the
disk will have different ``projection factors'' $\Gamma$ to the line
of sight.  Here $\Gamma$ is equivalent to the angle $\mu$ discussed
above.  We define a polar coordinate system centred at the centre of
the disk.  The angle $\theta$ is measured from the $x$-axis in the direction
of the positive $y$-axis.  For a given the orbital phase $\phi$, 
$\Gamma(r,\theta)$ is given by
\begin{eqnarray}
\Gamma(r,\theta) &=& -\cos\phi\sin i\sin\beta\cos\theta \nonumber \\
                 & & -\sin\phi\sin i\sin\beta\sin\theta \nonumber \\
                 & & +\cos i\cos\beta.
\end{eqnarray}
If $\Gamma(r,\theta)<0$, the point is not visible.

There are several geometrical details which we must account for 
when the disk is flared and/or when star 2 has a relatively large radius.
First, 
for cases when the  inclination $i$ is within $\beta$ degrees of 90,
parts of the disk face that have $\Gamma(r,\theta)>0$
will be below the rim as seen by a distant observer.
To account for these hidden points, we define the ``horizon'' of the top
rim of the disk.  In this discussion the horizon of an object is the outline
of the object in sky coordinates.
A point on the disk face is visible if its sky coordinates are {\em
inside} the top horizon.  Star 2 can block parts of the disk if its
radius is relatively large.  Since we currently require 
the inner radius of the disk to be equal
to the radius of star 2 (if present), some or all of the ``bottom'' part of
star 2 may be hidden by the disk.  Again, the shadowing of the disk by
star 2 and the lower part of star 2 by the disk is easily accounted for
by defining the suitable horizons.   Finally, a disk with a large radius
can inhibit
the mutual irradiation of the two stars since the ``top'' of star 1 cannot
``see'' the ``bottom'' of star 2, and vice-versa.   If a disk is present 
then inside the detailed reflection subroutine
each line of sight between points on star 1 and star 2 is checked
to see if it passes through the disk. 

Currently we assume that each surface element on the disk has a specific
intensity that is the same as a normal stellar atmosphere.  Following
Pringle et al.\ (\cite{pringle86}) 
we use the model with the largest
gravity for each effective temperature.  Of course, much more detailed
model atmospheres specific to accretion disks are available
(for example the grid of models for accretion disks in cataclysmic
variables presented by Wade \& Hubeny (\cite{wade98})), 
and in principle a separate
intensity table for accretion disks can almost trivially be added to ELC.
Indeed, 
Linnell \& Hubeny (\cite{linnell96}) 
create light curves for binaries with disks by first computing
detailed spectra for the disk using the Hubeny codes.  However, the
Hubeny code TLUSDISK is best suited for atmospheres hotter than about
10,000~K, and as such cannot really be used to model the cool outer parts
of many disks.   As it is now, our treatment of the accretion disk
is perhaps the most appropriate for systems where the disk is optically
faint and its main effect on the light curves is geometrical (i.e.\
it eclipses the bright mass-losing star).  Examples of such systems
are 
\object{W Crucis} (Zola \cite{zola96}),
\object{GRO J1655-40} (Orosz \& Bailyn \cite{orosz97}), and 
\object{BG Geminorum} (Benson et al.\ \cite{benson00}).

\subsection{Third light}

In many Algol type binaries there is good evidence for a fainter
third star that is gravitationally bound.  For example, the $O-C$
residuals of the eclipse timings of 
\object{SW Lyncis} are periodic and can be explained by the presence of a third
body in a 5.8 year eccentric orbit about the inner binary
(Ogloza et al.\ \cite{ogloza98}).  
In such triple systems, the ``third
light'' dilutes the observed amplitudes of the light curves from the binary,
provided of course that the third star is sufficiently bright.
We have a trivial way to self-consistently add third light 
to light curves in different bandpasses.  In ELC, 
the user specifies three parameters
for the third light:  the temperature of the third star, its gravity,
and its surface area relative to the surface area of star 1.  
The code interpolates the filter-integrated intensities for the third star
from the table, scales appropriately based on the surface area ratio, and
adds the light to each light curve.

\subsection{Accuracy of light curves through eclipse}

The integration of the observed flux from a single star is straightforward
and is sufficiently accurate for a reasonably small number of surface
elements.  However, quantisation errors can become noticeable in 
the light curve of a star going through an eclipse. 
In many cases (e.g.\ grazing eclipses, no reflection effect)
the number of surface elements can be modestly increased so that
a smooth light curve can be obtained without
a large increase in the CPU time required to compute the model.   
In other cases (e.g.\ deep eclipses and several iterations of
detailed reflection) the number of surface elements needed to get
smooth light curves becomes so large that the required CPU time becomes 
excessive.
Thus
ELC has two features which prove to
be quite effective in greatly reducing the numerical noise of light
curves though eclipse.  

\subsubsection{Improved horizon definition for the eclipsing star}  

The horizon of the
eclipsing star (the one ``in front'') is defined to be a collection of
points on the star which have $\mu=0$.
In previous versions of the code, the program would step though 
the surface grid
in the ``$\alpha$'' direction and record which surface elements were last
visible (i.e.\ the last point with $\mu>0$).  The resulting collection of
sky coordinates of these surface elements
would then define the horizon of the star.    
There is a systematic error introduced when the number of surface elements
is small since the numerical horizon of the star will be slightly smaller
than the actual horizon.
In the current version of the code,
the program steps along each latitude row and records the $\phi$-coordinate
of the last visible point $\phi_{\rm vis}$
and the $\phi$-coordinate of the
first point hidden below the horizon
$\phi_{\rm hid}$.
A simple bisection procedure is used to find the $\phi$-coordinate
(with a given latitude $\theta_i$) where $\mu=0$.  Fifteen iterations of
this bisection procedure are enough to find  $\phi_{\rm hor}$ to
better than $10^{-5}$ radians when $N_{\beta}=6$ (i.e.\ 24 longitude points).
The corresponding angles $\mu$ are all $\la 5\times
10^{-6}$.
The $x$ and $y$
sky coordinates of the point with the surface
coordinates $\theta_i,\phi_{\rm hid}$
are then determined.  A similar procedure is done 
where the program steps
through the $\theta$ angle for each longitude $\phi_i$ and $\theta_{\rm hor}$
is found from $\theta_{\rm vis}$ and $\theta_{\rm hid}$ using bisection.
After the list of points on the star with 
$\mu=0$ is generated, the $x$ and $y$ rectangular
coordinates of each point on the
sky are converted to a $R$,$\Theta$
polar coordinate system and sorted in the polar
angle $\Theta$.  
The sorted array forms a convex polygon on the sky.  If the number
of surface elements on the star is relatively small ($N_{\alpha}\la 30$
and $N_{\beta} \la 6$), then the actual horizon of the star can have
some curvature between adjacent points on the polygon.
Since the radius $R$ as a function of $\Theta$ is always
a very smooth function,  we use spline interpolation to resample
$R$ for every $1^{\circ}$ in $\Theta$.  The new resampled polygon with 360
points always has enough points so that the horizon of the star essentially
has no curvature between adjacent points.  We have done numerous tests
and found that the horizons derived using a small number of surface elements
with the new routine are always the same as the horizons found by the
old routine with a very large number of surface elements ($N_{\alpha}
\approx 400$ and $N_{\beta}\approx 100$).  

\subsubsection{Fractional surface elements on the eclipsed star}

To compute the observed flux in a given bandpass from a star at a
given phase, we numerically evaluate the integral in Eq.\ \ref{eq2}
using all of the surface elements with $\mu>0$:
\begin{equation}
I_{\rm FILT}=\sum_{i=1}^{N_{\alpha}}\sum_{j=1}^{4N_{\beta}}
I_{\rm FILT}(\mu_{i,j})\mu_{i,j}r^2_{i,j}\Delta\theta\Delta\phi/\cos\beta_{i,j}
\label{sumeq}
\end{equation}
where $\Delta\theta$ and $\Delta\phi$ are the angular spacings of the
elements in latitude and longitude, respectively, and
where $\beta$ is the angle between the surface normal and the radius from
the centre of the star.  In other words, we perform a simple numerical 
quadrature along each latitude row where the points are equally spaced
in the angle $\phi$.  
If there is an eclipsing body in front of the star whose flux
is being evaluated, then each point on the star in back
is projected onto the sky and a simple
routine is used to see if this point is inside the polygon
representing the horizon of the
body in front.  If the point in question is eclipsed, its flux contribution
is simply left out of the summation.  

In general, the horizon of the star in front will not pass exactly between
two adjacent points on a given latitude row on the star in back.  As a
result there will
be ``visible'' points  that are actually centred in partially eclipsed 
surface elements and
``eclipsed'' points that are centred in partially visible surface
elements.  If the number of surface elements on the star in back
is large enough, the contribution from the fractionally eclipsed surface
elements will tend to cancel out.  
However, it is much more computationally
efficient to use a smaller number of surface elements and make a correction
for the fractionally eclipsed pixels.  At each latitude row 
$\theta_i$ on the star
in back, the program determines the $\phi$-coordinate of the last
point visible
before the horizon of the eclipsing body $\phi_{\rm vis}$
and the 
$\phi$-coordinate of the first
point hidden
behind the horizon of the eclipsing body $\phi_{\rm hid}$.  Another
bisection procedure is used to determine the $\phi$-coordinate on the
star in back 
where the horizon of the eclipsing body intersects the 
$\theta_i$ latitude
row
$\phi_{\rm hor}$.  
If $|\phi_{\rm hor}-\phi_{\rm vis}|<\Delta\phi/2$, then the
last point visible before the horizon is centred in a partially eclipsed
surface element and a negative correction is added to the flux
summation:
\begin{equation}
F_{\rm corr}=\left[{2(\phi_{\rm hor}-\phi_{\rm vis})-\Delta\phi
\over \Delta\phi}\right]F_{\rm vis}
\end{equation}
where $F_{\rm vis}$ is flux from the last visible point, i.e.\
\begin{equation}
F_{\rm vis}=I_{\rm FILT}
(\mu_{i,j})\mu_{i,j}r^2_{i,j}\Delta\theta\Delta\phi/\cos\beta_{i,j}.
\end{equation}
Likewise, if $|\phi_{\rm hor}-\phi_{\rm hid}|<\Delta\phi/2$, then the
first point hidden behind the horizon is centred in a partially visible
surface element and a positive correction is added to the flux
summation:
\begin{equation}
F_{\rm corr}=\left[{\Delta\phi - 2(\phi_{\rm hor}-\phi_{\rm hid})
\over \Delta\phi}\right]F_{\rm hid}
\end{equation}
where $F_{\rm hid}$ is the flux the eclipsed point.
If there is an annular eclipse, then this procedure for finding
$\phi_{\rm hor}$ may have to be done twice for a given latitude row.
We have tested this simple interpolation procedure quite extensively and
have found it to be quite effective.  Smooth light curves through eclipse
can be obtained for grid sizes as small as $N_{\alpha}=14$
and $N_{\beta}=6$.   

\subsection{Comparison with Wilson and Devinney}

We have done extensive testing of the ELC code in its black body mode
against the W-D code.  Unfortunately, there is some confusion over the
notation of some of the input parameters between the two codes.  In
particular, $\Omega$ in ELC is the ratio of the star's angular velocity
to the orbital velocity, whereas $\Omega$ in W-D refers to the 
potential.  The W-D
$\Omega$-potentials essentially define the shapes of the
stars.
Perhaps to conserve parity, $f$ in ELC is the 
``filling factor'' which has the same function as the $\Omega$-potential
in W-D (i.e.\ it defines the shape of the star), 
whereas in W-D, $F$ is the ratio of the star's angular velocity
to the orbital velocity (ELC's $\Omega$).   The phase convention between
the two codes is different.  In ELC, star 1 is in front of star
2 at phase 0.0, whereas 
in W-D star 1 is {\em behind} star 2 at phase 0.0.

To facilitate comparisons between ELC and W-D, we do two things.  First,
since the
internal form of the potential is the same for both codes,  ELC prints
out the program values of the potentials which then are the input
$\Omega$ potentials for W-D.  Second, we add a phase shift of 0.5
to the W-D light curves (the 'pshift' input parameter).  We computed various
model binary light curves using the two codes and compared them by normalising
the light curves at phase 0.25 (quadrature).  We of course used
exactly the same effective wavelengths, limb darkening laws and coefficients,
and reflection schemes for both codes.  In most cases, the light curves
agreed to better than 0.1 per cent (better than 1 millimag).    If we
compared the light curves by adjusting the normalisation of one of 
the curves to match the other, then the largest deviations became even
smaller.  

ELC also computes radial  velocity curves 
(Wilson \& Sofia \cite{wilson76}). 
The velocity curves from ELC agree with the W-D velocity curves to better
than 0.1 per cent.

Finally, we compared computed geometric quantities between the two codes
(i.e.\ the ``polar'' radius, ``point'' radius, etc.).  The agreement was
essentially exact.  We also computed sphere-equivalent Roche lobe radii
and compared our results with Eggleton's 
(\cite{eggleton83}) results 
and likewise found nearly exact agreement.

\end{document}